\documentstyle[preprint,aps,epsfig]{revtex}
\begin{document}
\draft
%
%%%% title page %%% 
%
\preprint{$
\begin{array}{l} 
\mbox{KA-TP-26-2001}\\[-3mm]
\mbox{UB-HET-01-04}\\[-3mm]
\mbox{UR-1642}\\[-3mm]
\mbox{August 2001}\\[1cm]
\end{array}
$}
\title{Electroweak Radiative Corrections to Neutral-Current \\ Drell-Yan 
Processes at Hadron Colliders\\[1cm]}
\author{U.~Baur$^a$, O.~Brein$^b$, W.~Hollik$^b$, C.~Schappacher$^b$, 
and D.~Wackeroth$^c$\\[0.5cm]}
\address{$^a$Department of Physics,
State University of New York, Buffalo, NY 14260, USA\\
$^b$Institut f\"ur Theoretische Physik,
Universit\"at Karlsruhe, D-76128 Karlsruhe, Germany\\
$^c$Department of Physics \& Astronomy, University of Rochester,
Rochester, NY 14627, USA \\}
\date{\today}
\maketitle
%
%\tightenlines
%
%%% Abstract %%%
%
\begin{abstract}
\baselineskip14.pt %to keep abstract on 1 page 
We calculate the complete electroweak ${\cal O}(\alpha)$ corrections
to $p\,p\hskip-7pt\hbox{$^{^{(\!-\!)
}}$}  \to l^+ l^- X (l=e,\mu)$ in the Standard Model (SM) of electroweak
interactions.  They comprise weak and
photonic virtual one-loop corrections as well as real photon radiation
to the parton-level processes $q \bar q \to \gamma,Z \to l^+
l^-$. We study in detail the effect of the radiative corrections on the
$l^+l^-$ invariant mass distribution, the cross section in the $Z$ boson
resonance region, and on the forward-backward asymmetry, $A_{\rm FB}$, at the
Fermilab Tevatron and the CERN Large Hadron Collider. The 
weak corrections are found to increase the $Z$ boson cross
section by about 1\%, but have little effect on the forward-backward
asymmetry in the $Z$ peak region. Threshold effects of the 
$W$ box diagrams lead to pronounced effects in $A_{\rm FB}$ at 
$m(l^+l^-)\approx 160$~GeV which, however, will be difficult to observe 
experimentally. At high di-lepton invariant masses, the non-factorizable
weak corrections are found to become large. 
\end{abstract}
\newpage
\tightenlines
\section{Introduction}\label{sec:intro}
Drell-Yan production in hadronic collisions, $p\,p\hskip-7pt\hbox{$^{^{(\!-\!)
}}$} \to l^+l^-X$ 
($l=e,\,\mu$), is an interesting process for a number of reasons. 
Low mass Drell-Yan production is of interest because of the sensitivity
to parton distribution functions (PDFs) at small $x$
values~\cite{CDFDY,Abe:1999gq}. 
In the $Z$-boson resonance region, measurement of the $Z$-boson
mass, $M_Z$, and width, $\Gamma_Z$, and comparison with the values 
obtained at LEP helps to accurately 
calibrate detector components which is important for the determination of
the $W$ mass~\cite{cdfwmass,d0wmass,Brock:1999ep}. Measuring the
forward-backward 
asymmetry, $A_{\rm FB}$, in the vicinity of the $Z$ pole~\cite{AFBCDF} makes it
possible to extract the effective weak mixing angle. The ratio, $R$, 
of the $W\to l\nu$ and $Z\to l^+l^-$ cross sections can be used to extract
information on the width of the $W$ boson~\cite{D0Wcross,cdfr}. Finally,
above the $Z$ peak, one can search for physics beyond the SM such
as extra neutral gauge bosons~\cite{zprime}, effects of large extra
dimensions~\cite{xtradim1,xtradim2},
or composite quarks and leptons~\cite{comp} in either the di-lepton
invariant mass distribution or the forward-backward asymmetry. 

With the anticipated large data sets of $2-20~{\rm fb}^{-1}$ for Run~II
of the Fermilab Tevatron and 100~fb$^{-1}$ per year at the CERN Large 
Hadron Collider (LHC), it is very important to fully understand and
control higher order QCD and electroweak corrections to Drell-Yan 
production. A complete calculation of the full ${\cal O}(\alpha)$ 
radiative corrections to $p\,p\hskip-7pt\hbox{$^{^{(\!-\!)
}}$} \rightarrow \gamma,\, Z \rightarrow l^+ l^-$ has not
been carried out yet. In a previous calculation, only the QED
corrections had been included~\cite{Baur:1998wa} while ``genuine'' weak
corrections were ignored. At the Tevatron (LHC), the expected
statistical uncertainty on the $Z$ boson cross section for
2~fb$^{-1}$ (10~fb$^{-1}$) is approximately 0.2\% (0.05\%) per 
lepton channel. In contrast, the genuine weak corrections in the $Z$ 
peak region are expected to be of ${\cal O}(1\%)$
in magnitude and grow with the di-lepton invariant
mass, $m(l^+l^-)$, similar to di-fermion production in $e^+e^-$ 
collisions~\cite{cc}. It is thus necessary to include these
corrections when data and SM prediction are compared. Furthermore, in 
order to properly
calibrate the $Z$ boson mass and width using the available LEP data, it
is desirable to use exactly the same theoretical input which has been used to 
extract $M_Z$ and $\Gamma_Z$ at LEP, i.e. to include the reducible and
irreducible ${\cal O}(g^4m^2_t/M_W^2)$ corrections to the effective
leptonic weak mixing parameter, $\sin^2\theta_{\rm eff}^l$, and the $W$ mass, 
$M_W$~\cite{dg}, in the calculation. Here, $g$ is the $SU(2)_L$ 
coupling constant, and $m_t$ is the top quark mass. 

In this paper, we present a complete calculation of the electroweak ${\cal
O}(\alpha)$ corrections to $p\,p\hskip-7pt\hbox{$^{^{(\!-\!)
}}$} \rightarrow \gamma,\, Z \rightarrow l^+ l^-$ which also takes
into account the ${\cal O}(g^4m^2_t/M_W^2)$ corrections to
$\sin^2\theta_{\rm eff}^l$ and $M_W$. For the numerical evaluation, 
we use the Monte Carlo method for next-to-leading-order 
(NLO) calculations described in Ref.~\cite{NLOMC}. With the Monte Carlo 
method, it is easy to calculate a variety of observables simultaneously 
and to simulate detector response. The QED corrections
are taken from Ref.~\cite{Baur:1998wa} and the collinear 
singularities associated with initial state
photon radiation are removed by universal collinear counter terms
generated by ``renormalizing'' the parton distribution
functions~\cite{rujula,perlt,qedhs}, in complete analogy to gluon 
emission in QCD. Final state
charged lepton mass effects are included in our calculation in the 
following approximation. The lepton mass regularizes the collinear 
singularity associated with final state photon radiation. The associated
mass singular logarithms of the form $\ln(\hat s/m_l^2)$, where $\hat
s$ is the squared parton center of mass energy and $m_l$ is the
charged lepton mass, are included in our calculation, but the very small
terms of ${\cal O}(m_l^2/\hat s)$ are neglected. 

The technical details of our calculation are described in Sec.~II. The
electroweak corrections consist of the set of electroweak loop
contributions, including virtual photons, and of the emission of real
photons. To regularize the ultraviolet divergences associated with the
virtual corrections, we use dimensional regularization in the {\sc
on-shell} renormalization scheme~\cite{bo86}.  
After a brief summary of the calculation of the
QED corrections~\cite{Baur:1998wa}, analytical expressions for the
genuine weak corrections are presented. In the $Z$ pole region, the 
leading universal electroweak corrections can be expressed in terms of 
effective vector and axial vector couplings. These corrections can thus
be taken into account in form of an effective Born approximation (EBA)
where the tree level vector and axial vector coupling constants in the
expression of the Born cross section are replaced by the effective
vector and axial vector couplings. The remaining non-factorizable weak
corrections are small in the $Z$ pole region, but 
become important at high di-lepton invariant masses due to the presence 
of large Sudakov-like electroweak
logarithms of the form $\ln(m(l^+l^-)/M_V)$ ($V=W,\, Z$)~\cite{cc}. In 
Sec.~II, we also present a numerical comparison of the full ${\cal
O}(\alpha)$ cross section and the forward-backward asymmetry at parton level
with that obtained in the EBA and the Born approximation. Such a 
comparison is helpful to gain insight into how the weak corrections 
affect measurable quantities. 

Numerical results for $p\bar p$ collisions at $\sqrt{s}=2$~TeV and for
$pp$ collisions at $\sqrt{s}=14$~TeV are presented in Sec.~III. When 
${\cal O}(g^4m^2_t/M_W^2)$ corrections to $\sin^2\theta_{\rm eff}^l$ and
$M_W$ are taken into account, the weak corrections increase the
$Z$ boson cross section by about 1\% but have little effect on $A_{\rm FB}$
in the $Z$ pole region. Threshold effects of the 
$W$ box diagrams are found to lead to small but pronounced effects in
the forward-backward asymmetry at $m(l^+l^-)\approx 160$~GeV. At large
di-lepton invariant masses, the weak corrections reduce the
differential cross section by ${\cal O}(10\%)$. The effect on $A_{\rm FB}$
is somewhat smaller. Finally, our conclusions are presented in Sec.~IV. 
 
%
%%% Section II: Technical part %%%
%
\section{Electroweak radiative 
corrections to neutral current Drell-Yan processes}\label{sec:technical}
The parton-level differential Born cross section for charged lepton
pair production via photon and Z boson exchange in quark-antiquark
annihilation ($l=e,\mu$),
\begin{equation}
q(p)+\bar{q}(\bar p) \rightarrow \gamma,Z \to l^+(k_+)+ l^-(k_-),
\end{equation}
is given by
\begin{equation}\label{eq:xsec0}
 d {\hat \sigma^{(0)}}=  dP_{2f} \, \frac{1}{12} \sum 
 |A_{\gamma}^0+ A_Z^0|^2 (\hat s,\hat t,\hat u) \; ,
\end{equation}
where the sum is taken over the spin and color degrees of
freedom of the initial and final state fermions, and $dP_{2f}$ denotes
the two-particle phase space element.  The factor $1/12$ results from
summing and averaging over the quark spin and color degrees of freedom. 
The matrix elements
$A^0_{\gamma}$ and $A^0_Z$ describe the photon and Z boson exchange
processes at lowest order in perturbation theory.  In
terms of the kinematical variables of the parton system
\begin{equation}
 \hat{s} = (p + \bar{p})^2, \quad
  \hat t = (p-k_+)^2, \quad 
  \hat u = (p-k_-)^2,
\label{eq:mandel}
\end{equation}
the squared Born matrix elements for massless external fermions are
\begin{eqnarray}\label{eq:Zborn}
\sum |{\cal A}_{\gamma}^0|^2 & = & 8 \, Q_q^2 \, Q_l^2\, (4\pi \alpha)^2
\; \frac{(\hat t^2+\hat u^2)}{\hat s^2}~, \nonumber \\[1.mm]
\sum |{\cal A}_Z^0|^2 & = & 8 \, \frac{|\chi(\hat s)|^2}{\hat s^2} \, 
\left [(v_q^2+a_q^2)(v_l^2+a_l^2) (\hat t^2+\hat u^2) -  
4 v_q a_q v_l a_l\, (\hat t^2-\hat u^2) \right ]~, \nonumber \\[1.mm]
\sum 2 {\cal R}e ({\cal A}_Z^0 {\cal A}_{\gamma}^{0 *}) & = & 16 \, Q_q 
\, Q_l \, a_q \, a_l\, (4\pi \alpha) \,  
\left [ v_q v_l (\hat t^2+\hat u^2)-a_q a_l (\hat t^2-\hat u^2)\right ]
\, \frac{{\cal R}e  \chi(\hat{s})}{\hat s^2}~,
\end{eqnarray}
with
\begin{equation}\label{eq:coup}
v_f=\frac{1}{2 s_w c_w} (I_f^3-2 s_w^2 Q_f), \qquad  
a_f=\frac{I_f^3}{2 s_w c_w} \; ,
\end{equation}
parametrizing the $Zf \bar f$ ($f=l,q$) couplings. Here, 
$Q_f$ and $I_f^3$ denote the charge and third component of the weak isospin 
quantum numbers of the fermion~$f$, and 
$s_w \equiv\sin\theta_W, c_w \equiv \cos\theta_W$ with $\theta_W$ being the
weak mixing angle. $\alpha \equiv \alpha(0)$ is the electromagnetic fine
structure constant. The pole in the $Z$ boson propagator 
is regularized by assuming a complex $Z$ boson mass $M_c$
\begin{equation}\label{eq:zprop}
\chi(\hat s)=4 \pi \alpha \, \frac{\hat s}{(\hat s-M_c^2)} \; .
\end{equation}
In a perturbative calculation of the $Z$ propagator, a Dyson resummation
of one-particle-irreducible (1PI) (renormalized) $Z$ self energies is 
performed. The imaginary part of $M_c^2$ is related 
to the $Z$ decay width $\Gamma_Z$ by unitarity while the real part is given by
\begin{equation}
{\cal R}e M_c^2(\hat s=M_Z^2)=M_Z^2.
\end{equation}
In the {\sc on-shell} renormalization scheme~\cite{bo86} which we use,
the physical $Z$ boson mass $M_Z$ is equal to the renormalized mass. 

The Dyson resummation introduces the problem of defining the mass and 
width of the $Z$ boson, and a gauge invariant description of the 
scattering amplitude, order-by-order in perturbation theory.
As discussed in detail in Ref.~\cite{Wetzel:1983mh}, the ${\cal O}(\alpha^2)$ 
contributions to the imaginary part of $M_c^2$ must be taken into
account for a description of the $Z$ resonance at the one-loop level. 
A consistent expansion of ${\cal I}m M_c^2$ to ${\cal O}(\alpha^2)$
yields
\begin{equation} \label{eq:imags}
{\cal I}m M_c^2(\hat s)= {\cal I}m\left( \hat \Sigma^Z(\hat s) 
\left[1+{\cal R}e \hat \Pi^Z(M_Z^2)\right]+\hat \Sigma^Z_{(2)}(\hat s)-
\frac{(\hat \Sigma^{\gamma Z}(\hat s))^2}{\hat s + 
\hat \Sigma^{\gamma}(\hat s)}\right)~.
\end{equation}
$\hat \Sigma^Z$ and $\hat \Sigma^Z_{(2)}$ in Eq.~(\ref{eq:imags}) are 
the transverse parts of the
renormalized one and two-loop corrected 1PI Z self energies.
$\hat \Pi^Z$ (see Eq.~(\ref{eq:piz})) and $\hat \Sigma^{Z,\gamma,\gamma Z}$ 
(see Eqs.~(\ref{eq:rensigg}) --~(\ref{eq:rensiggz})) are the
(renormalized) self energy insertions into the $Z$ and photon
propagators.  The last term in Eq.~(\ref{eq:imags}) takes into account
that the photon and $Z$ boson do not propagate independently beyond
leading order in perturbation theory. 
The evaluation of ${\cal I}m M_c^2$ at $\hat s=M_Z^2$ corresponds to a
Laurent expansion around the complex pole and leads to a description
of the $Z$ resonance in terms of a constant width, i.e. $M_c^2=M_Z^2-i M_Z
\Gamma_Z^{(0+1)}$. At LEP energies the $\hat s$-dependence of the one and 
two loop contributions to the imaginary part of the 1PI Z self energy 
can be approximated by
\begin{equation}
{\cal I}m \hat\Sigma^Z(\hat s) = {\hat s\over M_Z^2}~{\cal I}m 
\hat\Sigma^Z(M_Z^2) \qquad 
{\cal I}m \hat\Sigma^Z_{(2)}(\hat s) = {\hat s\over M_Z^2}~{\cal I}m 
\hat\Sigma^Z_{(2)}(M_Z^2) 
\end{equation}
so that $M_c^2=M_Z^2-i (\hat s/M_Z)
\Gamma_Z^{(0+1)}$.  Both descriptions are related by a transformation
of the parameters of the $Z$ resonance, $M_Z$ and $\Gamma_Z$, and the
residue of the complex pole, and are thus equivalent~\cite{Bardin:1988xt}.
In the following, we use the $\hat s$-dependent width approach.  The 
one-loop corrected $Z$ boson decay width, $\Gamma_Z^{(0+1)}$, is
discussed in Appendix~\ref{app:zdecay}.
 
The electroweak ${\cal O}(\alpha)$ corrections to neutral-current
Drell-Yan processes naturally decompose into QED and weak
contributions, i.e. they form gauge invariant subsets, and thus can
be discussed separately.  The observable NLO cross section is obtained
by convoluting the parton cross section with the quark distribution
functions $f_{q/A} (x,Q^2)$ ($\hat s=x_1 x_2 s$) and summing over all
quark flavors $q$ 
\begin{eqnarray}\label{eq:xsecobs}
d\sigma(s) &  =  & \sum_q \int_0^1 d x_1 d x_2 \, \bigg(  f_{q/A}
(x_1,Q^2) f_{\bar{q}/B} (x_2,Q^2) + (q \leftrightarrow \bar{q})  \bigg)
%q(x_1,Q^2) \, \bar q(x_2,Q^2) 
\left [d\hat \sigma^{(0+1)}(\hat s,\hat t,\hat u) \right
. \\[1.mm] \nonumber
&  & + \left . d\hat \sigma_{{\rm QED}}(\mu_{{\rm 
QED}}^2,\hat s,\hat t,\hat u)\right ]~,
\end{eqnarray}
with $(A,B)=(p,\bar{p})$ for the Tevatron and $(p,p)$ for the LHC.
$d\hat \sigma^{(0+1)}$ comprises the NLO cross section
including the weak corrections and $d\hat \sigma_{{\rm QED}}$
describes the QED part, i.e. virtual and real photon emission off the
quarks and charged leptons. The parton distribution functions 
depend on the QCD renormalization and factorization scales, $\mu_r$ and 
$\mu_f$, which we choose to be equal, $\mu_r=\mu_f=Q$.
The radiation of collinear photons off quarks requires the
factorization of the arising mass singularities into the PDFs which
introduces a QED factorization scale $\mu_{\rm QED}$ as will be explained
in more detail in the next section.
\subsection{QED corrections}\label{sec:qed}
QED radiative corrections consist of the emission of real and virtual
photons off the quarks and charged leptons.  The ${\cal O}(\alpha)$
QED corrections to $q \bar q\to \gamma ,Z\to l^+ l^- $ can be further
divided into gauge invariant subsets corresponding to initial and
final-state radiation. Since the incoming quarks are assumed to be massless 
in the parton model, initial state photon radiation results in collinear
singularities. These singularities are universal to all orders in 
perturbation theory and can be absorbed by a redefinition ({\it
renormalization}) of the 
PDFs~\cite{rujula}. This can be done in complete analogy to the
calculation of QCD radiative corrections.  As a result, 
the renormalized parton distribution functions become dependent on the QED
factorization scale $\mu_{\rm QED}$ which is controlled by the well-known
Gribov-Lipatov-Altarelli-Parisi (GLAP) equations \cite{ap}. These
universal photonic corrections can be taken into account by a
straightforward modification \cite{perlt,qedhs} of the standard GLAP
equations which describe gluonic corrections only. The modification
consists of an additional term which is proportional to the
electromagnetic fine-structure constant, $\alpha$, resulting in
modified distribution functions $q_f(x,\mu_{\rm QED}^2)$ for quarks with
flavor~$f$.  The gluon distribution $g(x,\mu_{\rm QED}^2)$ is only affected 
indirectly by QED corrections through terms of 
${\cal O}(\alpha \alpha_s)$.  The QED induced terms in the GLAP
equations lead to small, negative corrections at the per-mille level 
to the distribution functions for most values of $x$ and
$\mu_{\rm QED}^2$~\cite{Haywood:1999qg}. Only
at large $x \ {\stackrel{>}{\scriptstyle \sim}} \ 0.5$ and large
$\mu_{\rm QED}^2 \ {\stackrel{>}{\scriptstyle \sim}} \ 10^3$~GeV$^2$ do
the corrections reach the magnitude of one per cent. 

In order to treat the ${\cal O}(\alpha)$ initial state QED corrections
to Drell-Yan production in hadronic collisions in a consistent way, QED
corrections should be incorporated in the global fitting of the PDFs,
i.e. all data which are used to fit the parton distribution functions
should be corrected for QED effects. Current fits~\cite{mrst99,cteq5} to
the PDFs do not include QED corrections. The missing QED corrections 
introduce an
uncertainty which, however, is probably much smaller than the present
experimental uncertainties on the parton distribution functions. 

Absorbing the collinear singularity into the PDFs introduces a
QED factorization scheme dependence. The squared matrix elements for different
QED factorization schemes differ by the finite ${\cal O}(\alpha)$ terms
which are absorbed into the PDFs in addition to the singular terms. 

For our calculation, we have taken the ${\cal O}(\alpha)$ QED corrections
from Ref.~\cite{Baur:1998wa}. The calculation presented in
Ref.~\cite{Baur:1998wa} is based on an explicit diagrammatic
approach. The collinear singularities associated
with initial-state photon radiation are factorized into the parton
distribution functions as described above, but QED corrections are not
taken into account into the GLAP evolution of the PDFs.  
The ${\cal O}(\alpha)$ QED corrections are implemented both in the 
QED ${\rm \overline{MS}}$ and DIS schemes, which are defined
analogously to the usual ${\rm \overline{MS}}$~\cite{MSBAR} and 
DIS~\cite{OWENSTUNG} schemes used in QCD calculations. All numerical
calculations in this paper are performed using the QED DIS scheme.
The QED DIS scheme is defined
by requiring the same expression for the leading and next-to-leading order 
structure function $F_2$ in deep inelastic scattering. Since $F_2$ data
are an important ingredient in extracting PDFs, the effect of
the ${\cal O}(\alpha)$ QED corrections on the PDFs should be reduced
in the QED DIS scheme.
The collinear singularities associated with photon radiation
from the final state lepton lines are regulated by finite lepton
masses. 

\subsection{Non-QED corrections and the effective Born approximation} 
\label{sec:nonqed}
\begin{figure}
\phantom{x}
\vskip 1.cm
\centerline{\epsfig{figure=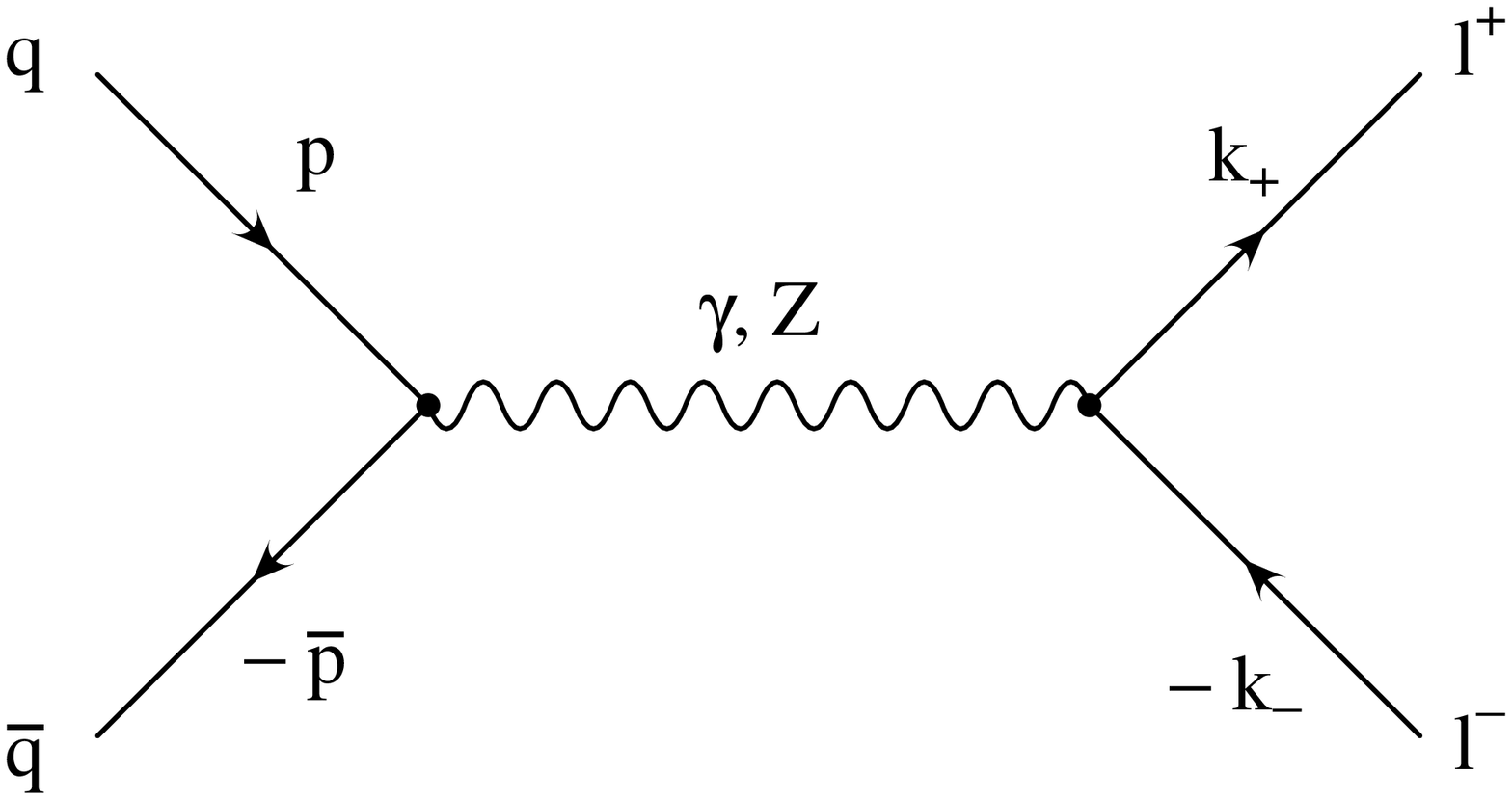,width=5cm}}
\vskip 9.mm
\centerline{\epsfig{figure=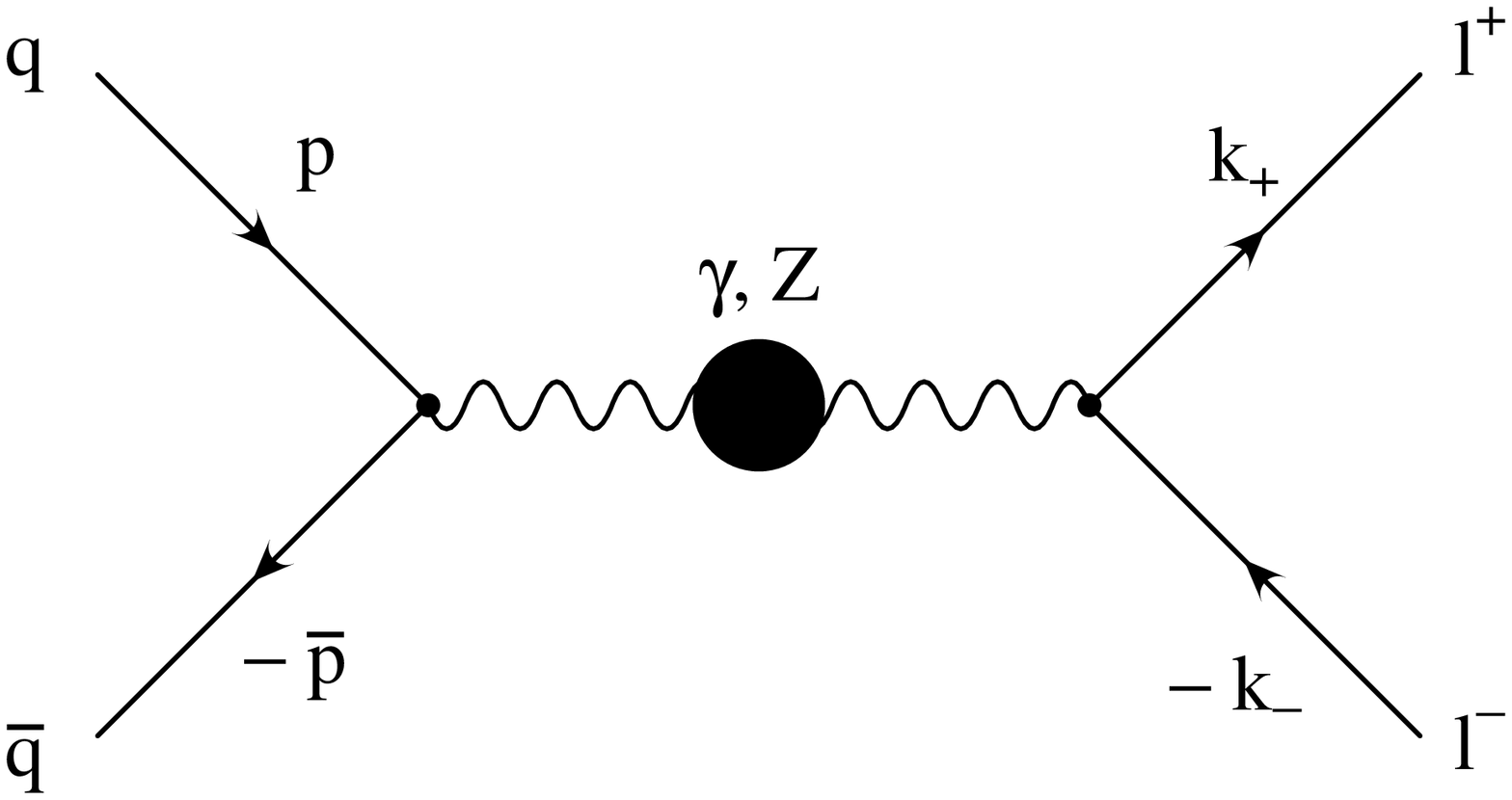,width=5cm} 
\epsfig{figure=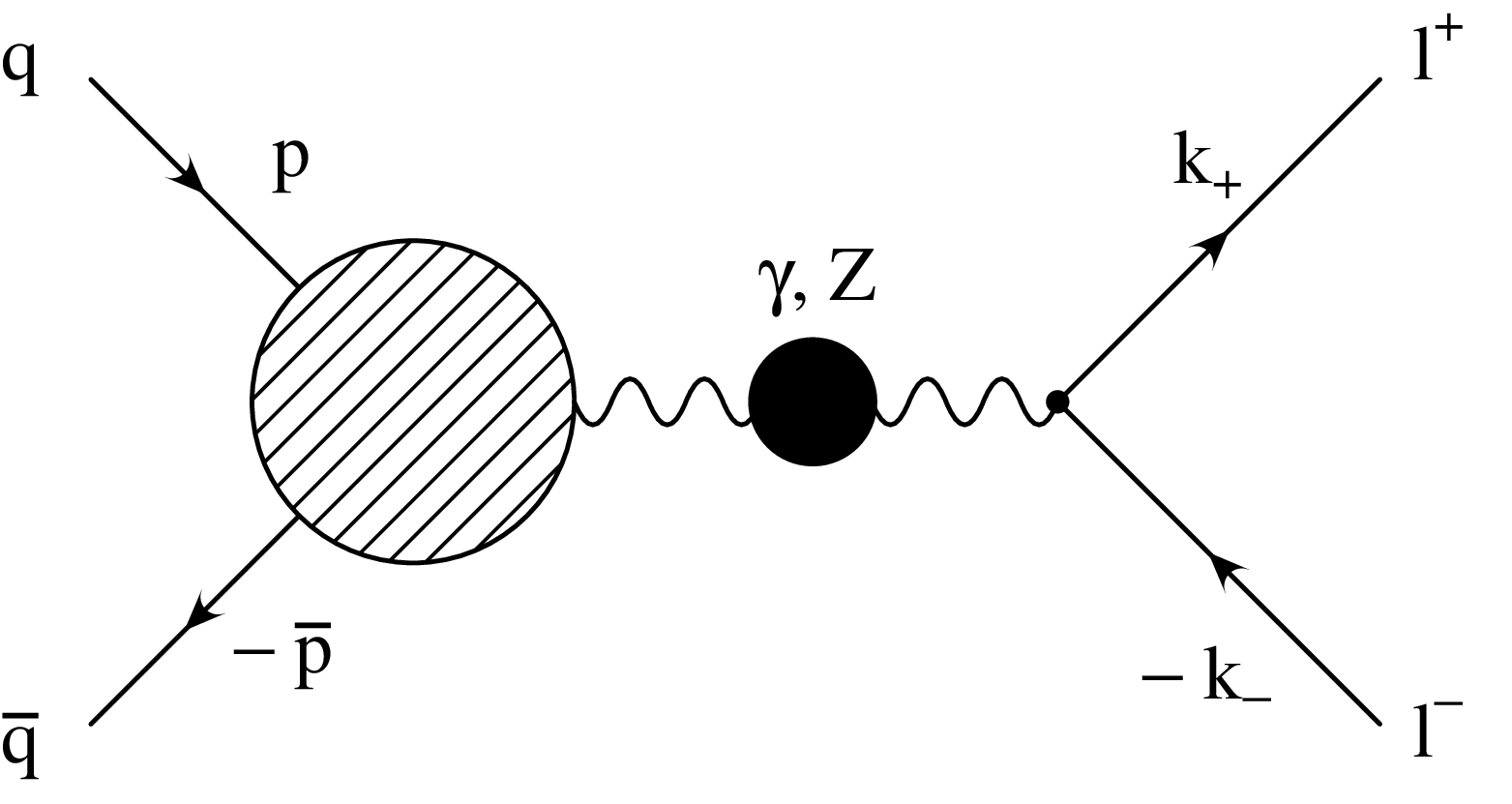,width=5cm}
\epsfig{figure=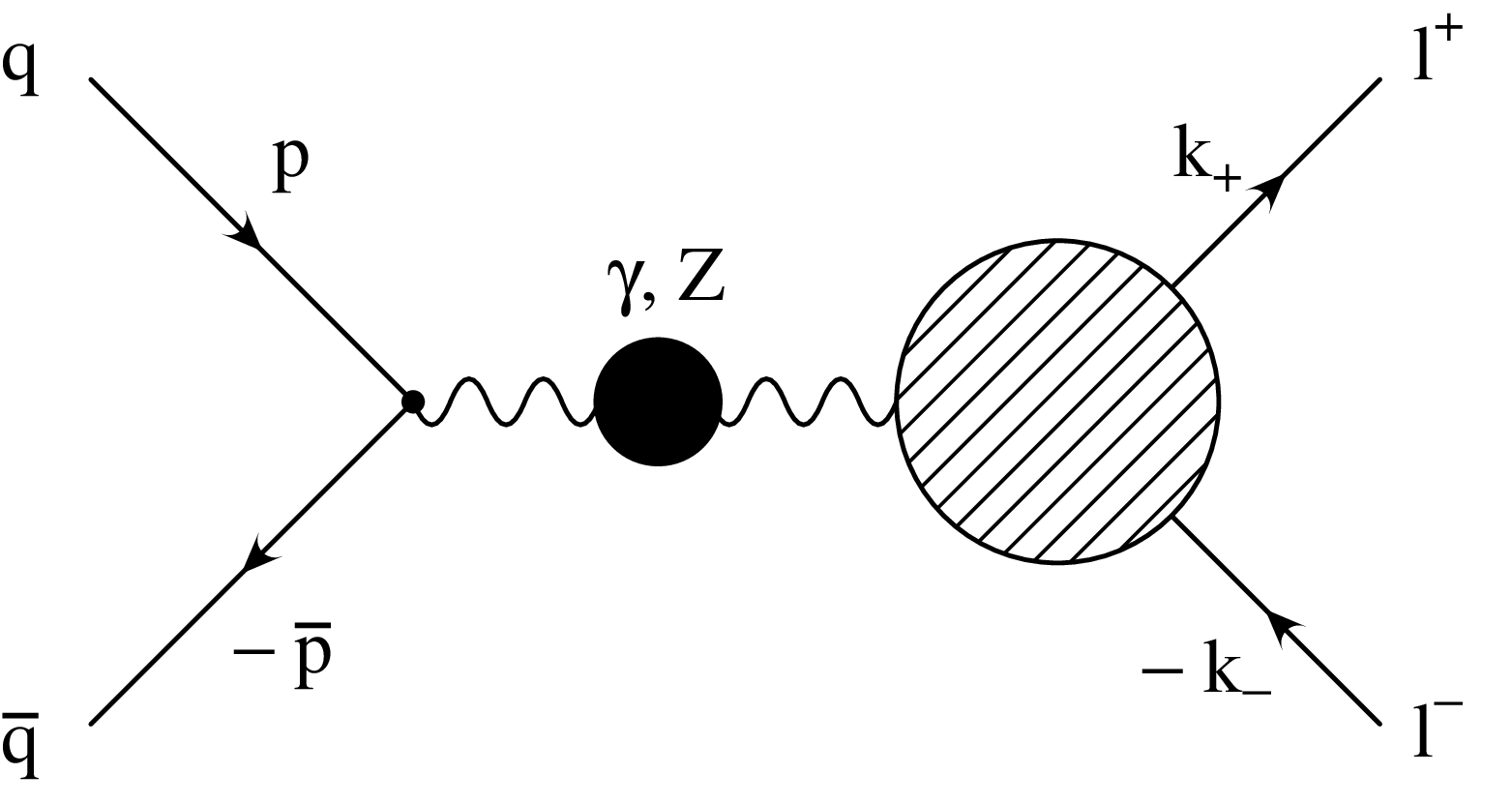,width=5cm}}
\vskip 9.mm
\caption{Born and higher-order weak contributions to $q\bar{q} 
\rightarrow \gamma,Z \to l^+ l^-$ in symbolic notation. The dark blob 
indicates the inclusion of all 1PI contributions to the photon and $Z$
propagators.}
\label{fig:zdiagrams}
\end{figure}

\begin{figure}
\centerline{\epsfig{figure=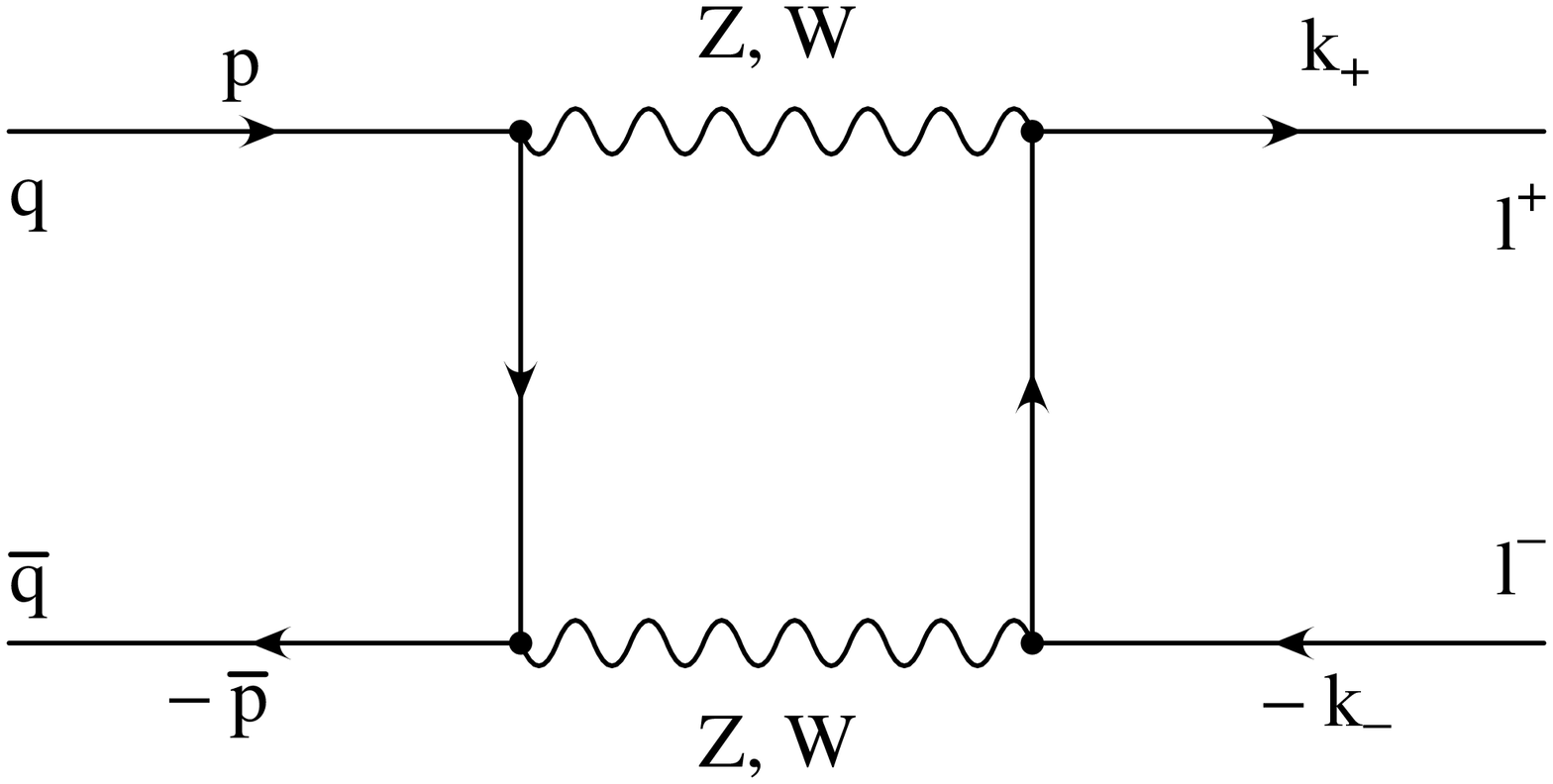,width=5cm} 
\hspace{1.5cm} \epsfig{figure=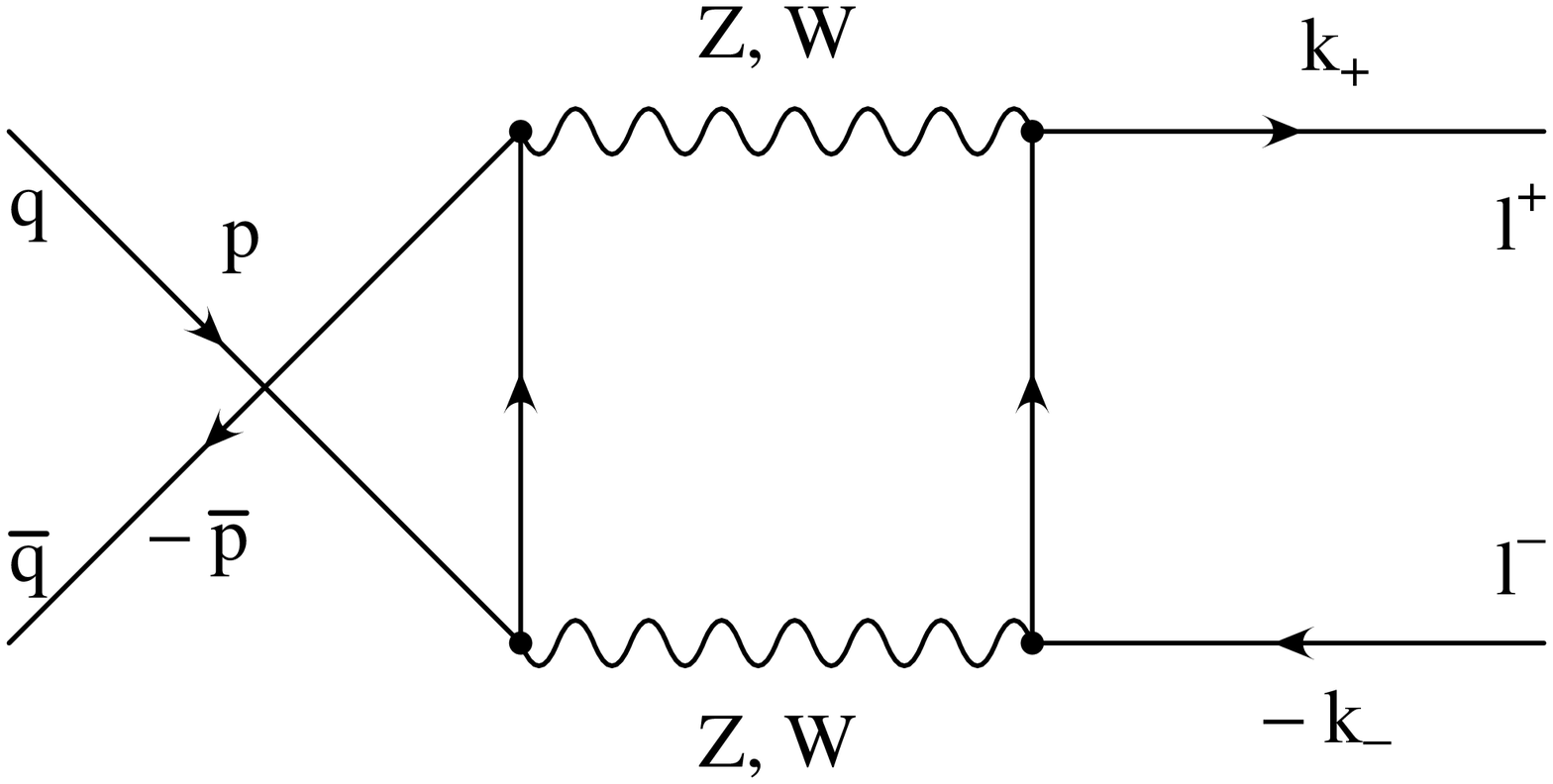,width=5.2cm}}
\vskip 9.mm
\caption[]{Box diagrams contributing to $q\bar{q} \to l^+ l^-$.}
\label{fig:box} 
\end{figure}

The non-QED corrections consist of self-energy contributions
to the photon and $Z$ propagators, vertex corrections to the
$\gamma/Z$-$l^+l^-$ and $\gamma/Z$-$q\bar{q}$ couplings (see 
Fig.~\ref{fig:zdiagrams}), and box
diagrams with two massive gauge bosons (see Fig.~\ref{fig:box}). 
Since we neglect all non-logarithmic fermion mass terms, there is no
Higgs boson contribution to the box diagrams and vertex corrections.
The calculation of the radiative corrections is performed in the
't~Hooft-Feynman gauge. To regularize and remove the arising ultraviolet
(UV) divergences we use dimensional regularization in the {\sc on-shell}
renormalization scheme as described in Ref.~\cite{bo86}.

In the following we closely follow Refs.~\cite{ho90} 
and~\cite{yellowbook95}, in particular for the treatment of additional 
higher-order corrections. For an accurate description of the $Z$
resonance, it is important to take these corrections into account. 
The NLO differential cross section at the parton level 
including weak ${\cal O}(\alpha)$ and leading ${\cal O}(\alpha^2)$
corrections is of the form
\begin{equation}\label{eq:xsecweak}
d \hat \sigma^{(0+1)}= dP_{2f} \, \frac{1}{12} \, \sum 
\left |A_{\gamma}^{(0+1)}+ A_Z^{(0+1)}\right |^2(\hat s,\hat t,\hat u) + 
d\hat \sigma_{{\rm box}}(\hat s,\hat t,\hat u) \; .
\end{equation} 
Here, $d\hat \sigma_{{\rm box}}$ describes
the contribution of the box diagrams shown in Fig.~\ref{fig:box}. 
The matrix elements
$A_{\gamma,Z}^{(0+1)}$ comprise the Born matrix elements, $A_{\gamma,Z}^0$, 
the $\gamma,Z$ and $\gamma Z$ self energy insertions, 
including a leading log resummation of the terms involving the light
fermions, and the one-loop vertex corrections, as shown in 
Fig.~\ref{fig:zdiagrams}.
$A_{\gamma,Z}^{(0+1)}$ can be expressed in terms of effective
vector and axial-vector couplings, $g_{V,A}^{(\gamma,Z),f}$ ($f=l,q$), 
such that the squared matrix elements for massless external fermions 
can be expressed in the form
\begin{eqnarray}\label{eq:ampweak}
\sum |A_{\gamma}^{(0+1)}|^2 &= & \frac{(4\pi\alpha)^2}{[1+{\cal R}e 
\hat \Pi^\gamma(\hat s)]^2 \,  \hat s^2} \times 
\nonumber \\[3.mm]
& & 8\,\Bigl[(|g_V^{\gamma,l}|^2+|g_A^{\gamma,l}|^2) \,
(|g_V^{\gamma,q}|^2+|g_A^{\gamma,q}|^2) \, (\hat t^2+\hat u^2) 
\nonumber \\[1.mm]
&-& 4\,{\cal R}e (g_V^{\gamma,l} (g_A^{\gamma,l})^{*}) \, 
{\cal R}e (g_V^{\gamma,q} (g_A^{\gamma,q})^{*}) \, (\hat t^2-\hat u^2)
\Bigr]~, \\[5.mm]
\sum |A_{Z}^{(0+1)}|^2 &= & \frac{|\chi(\hat s)|^2}{[1+{\cal R}e \hat 
\Pi^Z(\hat s)]^2 \,  \hat s^2} \times 
\nonumber \\[3.mm]
& & 8\,\Bigl[(|g_V^{Z,l}|^2+|g_A^{Z,l}|^2) \,
(|g_V^{Z,q}|^2+|g_A^{Z,q}|^2) \, (\hat t^2+\hat u^2) 
\nonumber \\[1.mm]
&-& 4\,{\cal R}e (g_V^{Z,l} (g_A^{Z,l})^{*}) \, 
{\cal R}e (g_V^{Z,q} (g_A^{Z,q})^{*}) \, (\hat t^2-\hat u^2) \Bigr]~,
\\[5.mm]
\sum 2 {\cal R}e (A_Z^{(0+1)} A_{\gamma}^{(0+1) *}) &= & 
\frac{(4\pi\alpha) \, |\chi(\hat s)|^2}{[1+{\cal R}e \hat
\Pi^{\gamma}(\hat s)] 
 \,  [1+{\cal R}e \hat \Pi^Z(\hat s)] \,  \hat s^2} \, 16\,{\cal R}e \,  
\Bigl( \chi^{-1}(\hat s)\times 
\nonumber \\[3.mm]
& &  \Bigl[ 
 (g_V^{Z,l} (g_V^{\gamma,l})^{*}+ g_A^{Z,l} (g_A^{\gamma,l})^{*}) 
(g_V^{Z,q} (g_V^{\gamma,q})^{*}+ g_A^{Z,q} (g_A^{\gamma,q})^{*}) (\hat
t^2+ \hat u^2)
\nonumber \\[1.mm]
&- & (g_A^{Z,l} (g_V^{\gamma,l})^{*}+g_V^{Z,l} (g_A^{\gamma,l})^{*}) 
(g_A^{Z,q} (g_V^{\gamma,q})^{*}+g_V^{Z,q}\,(g_A^{\gamma,q})^{*}) (\hat
t^2-\hat u^2)\Bigr] \Bigl)~,
\end{eqnarray}
with
\begin{eqnarray}\label{eq:effcoup}
g_A^{Z,f}(\hat s)& =& a_f+G_A^{Z,f}(\hat s)\,,
\nonumber \\
g_V^{Z,f}(\hat s)& =& v_f+F_V^{Z,f}(\hat s)+Q_f 
\frac{\hat \Pi^{\gamma Z}(\hat s)}{1+\hat \Pi^{\gamma}(\hat s)}\,,
\nonumber \\
g_A^{\gamma,f}(\hat s)& =& -G_A^{\gamma,f}(\hat s) \,,
\nonumber \\
g_V^{\gamma,f}(\hat s)& =& Q_f-F_V^{\gamma,f}(\hat s)  \; .
\end{eqnarray}
$F_V^{(\gamma,Z),f}$ and $G_A^{(\gamma,Z),f}$ denote the renormalized 
vector and axial-vector formfactors which parametrize the weak 
corrections to the $(\gamma,Z) f \bar f$ vertices. $\hat \Pi^X, 
X=\gamma, Z, \gamma Z$, describe the renormalized photon, $Z$ and $(\gamma,Z)$ 
self energy insertions, 
\begin{eqnarray}\label{eq:pigamma}
\hat \Pi^\gamma(\hat s)&=&\frac{\hat \Sigma^\gamma(\hat s)}{\hat s}~,
\\[2.mm]
\label{eq:pigamz}
\hat \Pi^{\gamma Z}(\hat s)&=&\frac{\hat \Sigma^{\gamma Z}(\hat s)}{\hat s}~,
\\[2.mm]
\label{eq:piz}
\hat \Pi^Z(\hat s)&=&\frac{1}{\hat s-M_Z^2} \left(\hat \Sigma^{Z}(\hat s)-
\frac{(\hat \Sigma^{\gamma Z}(\hat s))^2}{\hat s+\hat \Sigma^\gamma(\hat
s)}\right) \; .
\end{eqnarray}
The box contribution $d\hat \sigma_{{\rm box}}$ 
cannot be absorbed in effective couplings. However, 
in the $Z$ resonance region, the box diagrams can be
neglected and the NLO cross section $d\hat \sigma^{(0+1)}$ 
is of Born-structure.
For more details about the self energies, form factors and box contributions
we refer to Appendices~\ref{app:sv} and~\ref{app:box}, respectively.
In Appendix~\ref{app:sv} we also describe 
the inclusion of leading higher-order (irreducible) QCD and electroweak 
corrections connected to the $\rho$ parameter.

In the $Z$ resonance region, the dominant contributions to 
the non-photonic electroweak corrections can be taken into account by 
redefining several quantities appearing in the expression of 
the Born cross section given in Eq.~(\ref{eq:xsec0}). In the resulting 
effective Born approximation (EBA), the fine structure constant,
$\alpha$, is replaced by the running electromagnetic coupling,
$\alpha(\hat s)$
\begin{equation}\label{eq:alphaz}
 \alpha \to  \alpha(\hat s)=\frac{\alpha}{1-\Delta\alpha(\hat{s})}\, , \quad
\Delta \alpha(\hat s)= - {\cal R}e \hat \Pi^{\gamma}_{{\rm ferm}}(\hat s) \; ,
\end{equation}
where $\hat \Pi_{ferm}^{\gamma}$ denotes the fermion-loop contribution
to the photon vacuum polarization. $\chi(\hat s)$ is expressed in terms
of the physical $W$ and $Z$ boson masses, the $Z$ width measured at LEP,
and the Fermi constant $G_\mu$,
\begin{equation}\label{eq:zpropgmu}
 \chi(\hat{s}) = 4 \sqrt{2} G_\mu M_W^2 s_w^2\, 
                 \frac{\hat{s}}{\hat{s}-M_Z^2 + 
                  {\rm i}\hat{s} \Gamma_Z/M_Z } \, ,
\end{equation} 
where
\begin{equation}
s_w^2=\left(1-{M_W^2\over M_Z^2}\right ).
\end{equation}
Finally, the vertex and self energy corrections are taken into account by
replacing 
\begin{equation}\label{eq:veff} 
v_f \to v_f^{{\rm eff}} = {1\over 2s_w}~{M_Z\over M_W}~(I_3^f -2 Q_f
\sin^2\theta_{\rm eff}^f) \, , 
\quad f=l,q \; ,
\end{equation}
where 
\begin{equation}\label{eq:seff}
\sin^2\theta_{\rm eff}^f = \frac{1}{4 |Q_f|} \; 
\left(1-\frac{{\cal R}e g_V^{Z,f}(M_Z^2)}{{\cal R}e g_A^{Z,f}(M_Z^2) }
\right)~,
\end{equation}
is the effective electroweak mixing parameter for leptons at the $Z$ peak. 
$\sin^2\theta_{\rm eff}^l$ has been measured at LEP and the SLC. The 
effective weak mixing parameter for up and down type quarks is 
approximately given by
\begin{eqnarray}
\sin^2\theta_{\rm eff}^u & \approx & \sin^2\theta_{\rm eff}^l-0.0001,\\
\sin^2\theta_{\rm eff}^d & \approx & \sin^2\theta_{\rm eff}^l-0.0002.
\end{eqnarray}

Above the $Z$ peak region, the effective Born approximation becomes
insufficient for two reasons: the effective couplings are not static
but grow as functions of $\hat{s}$, and the box diagrams are no longer
negligible. Their contribution increases strongly with energy and they 
contribute significantly at high invariant masses of the lepton pair.

\subsection{Numerical discussion of the non-QED
corrections at the parton level}
\label{sec:parton}
Before we present results at the hadron level, we discuss the 
observables of interest at the parton
level, where many characteristics of the weak corrections manifest
themselves. For the numerical evaluation we chose the
following set of SM input parameters~\cite{caso}:
\begin{eqnarray}\label{eq:pars}
G_{\mu} = 1.16639\times 10^{-5} \; {\rm GeV}^{-2}, 
& \qquad & \alpha= 1/137.0359895, 
\nonumber \\ 
M_Z = 91.1867 \; {\rm GeV}, & \quad & 
\alpha_s\equiv\alpha_s(M_Z^2)=0.119, 
\nonumber  \\
m_e  = 0.51099907 \; {\rm MeV}, &\quad &m_{\mu}=0.105658389 \; {\rm GeV},  
\quad m_{\tau}=1.777 \; {\rm GeV},
\nonumber  \\
m_u=0.0464 \; {\rm GeV}, & \quad & m_c=1.5 \; {\rm GeV}, 
\quad m_t=174 \; {\rm GeV},
\nonumber  \\
m_d=0.0465 \; {\rm GeV}, & \quad & m_s=0.15 \; {\rm GeV}, \quad m_b=4.7
\; {\rm GeV}. 
\end{eqnarray}
The fermion
masses only enter through loop contributions to the vector boson self
energies and as regulators of the collinear singularities which arise
in the calculation of the QED contribution. Non-zero light quark masses 
are only used in the calculation of the vector boson self energies. The 
light quark masses are
chosen such that the value for the hadronic contribution to the photon 
vacuum polarization for five active flavors, $\Delta
\alpha_{had}^{(5)}(M_Z^2)=0.028$~\cite{Eidelman:1995ny},
which is derived from low-energy $e^+ e^-$ data with the help of
dispersion relations, is recovered.  

The $W$ mass and the Higgs boson mass, $M_H$, are related via loop 
corrections. A parametrization of the $W$ mass which, for 
$65~{\rm GeV}<M_H<1$~TeV, deviates by at most
0.4~MeV from the theoretical value including the full fermionic two-loop
contributions is given in Ref.~\cite{freitas}. Here we use the
parametrization of Ref.~\cite{Degrassi:1998iy} 
\begin{eqnarray}
\label{eq:param}
M_W&=&M_W^0-0.0581 \; \ln\left(\frac{M_H}{100~{\rm GeV}}\right)
-0.0078 \ln^2\left(\frac{M_H}{100~{\rm GeV}}\right)
-0.085 \; \left(\frac{\alpha_s}{0.118}-1\right)
\nonumber \\[2.mm]
&-& 0.518 \; \left(\frac{\Delta \alpha_{had}^{(5)}(M_Z^2)}{0.028}-1\right)
+ 0.537 \; \left(\Bigl(\frac{m_t}{175~{\rm GeV}}\Bigr)^2-1\right)
\end{eqnarray}
with $M_W^0=80.3805$ GeV, which was used in the
analysis of the LEP data.  The parametrization of Eq.~(\ref{eq:param})
reproduces the result of Ref.~\cite{Degrassi:1998iy} to 0.2~MeV for 
$75~{\rm GeV}<M_H<350$~GeV. For the numerical discussion we choose
\begin{equation}
\label{eq:mhiggs}
M_H=120~{\rm GeV}, 
\end{equation}
which is consistent with current direct~\cite{hbound1} and indirect
bounds~\cite{hbound2}, and work in the $s$-dependent width scheme. 
The $Z$-boson decay width is calculated including electroweak and QCD
corrections as described in Appendix~\ref{app:zdecay}.  The NLO
prediction for the $Z$ boson width is also used in the
calculation of the lowest-order and EBA predictions.  
For the input parameters listed in Eq.~(\ref{eq:pars}) we obtain for
the effective leptonic weak mixing angle of Eq.~(\ref{eq:seff}) $\sin^2
\theta^l_{{\rm eff}}=0.23167$, and
$\Gamma_Z=\Gamma_Z^{(0+1)}=2.4932$~GeV for the width of the $Z$ boson. 

\begin{figure}
\phantom{x}
\vskip 16.cm
\includegraphics{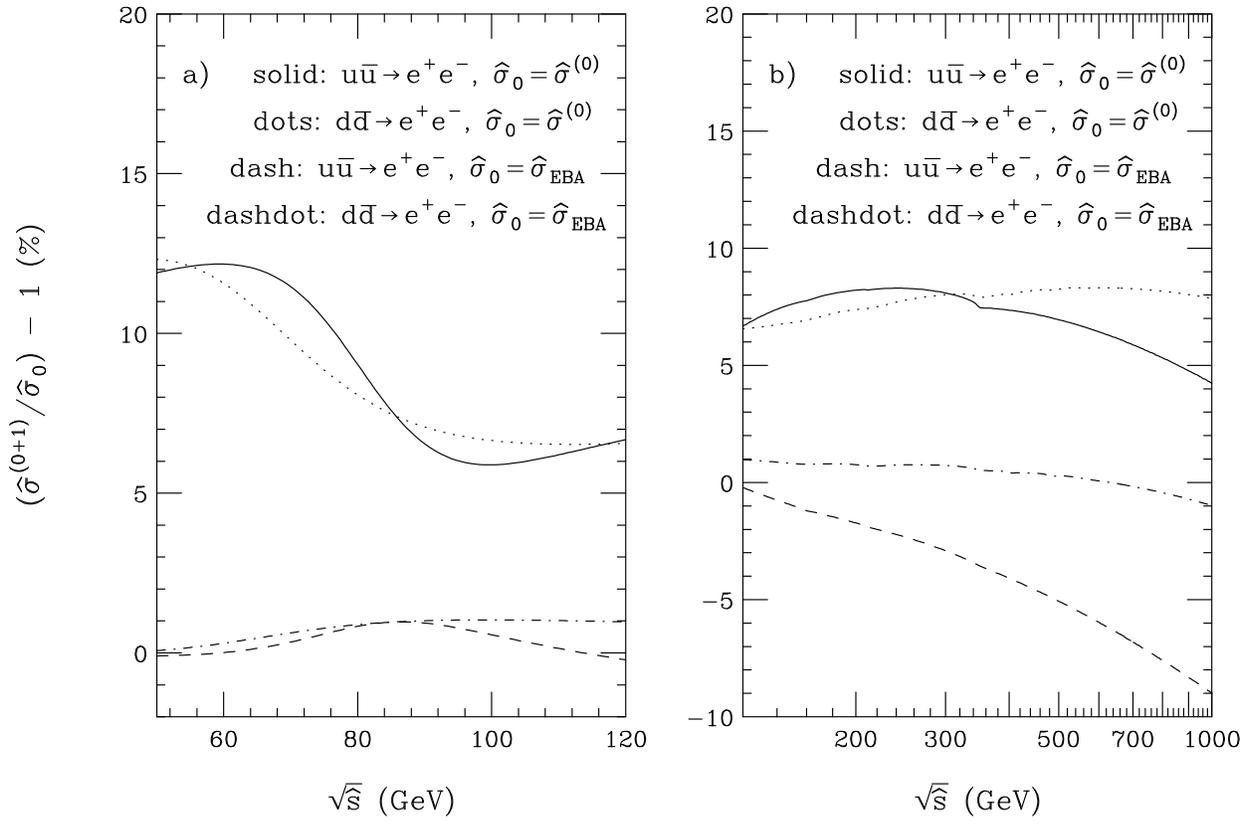}
\caption{The relative corrections to
the total cross sections for $u \bar u \to e^+ e^-$ and 
$d \bar d \to e^+ e^-$ a) in the vicinity of the
$Z$ resonance, and b) at high parton center of mass energies.}
\label{fig:z1}
\end{figure}

In the following we discuss the impact of the weak
corrections on the total cross sections and the forward-backward
asymmetries for $u \bar u~(d \bar d)\to \gamma,Z \to e^+ e^-$ as a
function of the parton center of mass energy $\sqrt{\hat s}$. Almost
identical results are obtained for the $\mu^+\mu^-$ final state. We
compare the full NLO result, $d\hat \sigma^{(0+1)}$ of
Eq.~(\ref{eq:xsecweak}), with the Born prediction, $d\hat
\sigma^{(0)}$ of Eq.~(\ref{eq:xsec0}), and the result obtained in the 
EBA. In Fig.~\ref{fig:z1}, we show the relative corrections
$\hat\sigma^{(0+1)}/\hat\sigma_0-1$, in per cent, to the total cross 
sections for the processes $u\bar{u} \to e^+e^-$ and
$d\bar{d} \to e^+e^-$ in the $Z$ resonance region (Fig.~\ref{fig:z1}a),
and at high parton center of mass energies (Fig.~\ref{fig:z1}b).
$\hat\sigma_0$ is either taken to be the Born cross section,
$\hat\sigma^{(0)}$, 
or the EBA prediction for the total cross section, $\hat\sigma_{{\rm EBA}}$. 
The weak corrections are seen to enhance the total cross
section by $10-12\%$ below, and by $5-7\%$ at and above the $Z$ peak.
The kink at $\sqrt{\hat s}\approx 350$~GeV in the solid and dotted lines
is due to the top quark threshold in the running coupling, $\alpha(\hat
s)$. Since this enhancement is mainly caused by universal electroweak
corrections, i.e. the running of $\alpha$ and corrections connected
to the $\rho$ parameter, the EBA represents a good description of the
NLO result in the $Z$ resonance region
($\hat\sigma^{(0+1)}/\hat\sigma_0-1\leq 1\%$). The difference between
the full NLO result and that obtained in the EBA is a measure for the 
effects of the non-universal corrections. At higher parton center 
of mass energies
the strong deviation of the EBA cross section from the NLO result is
essentially due to the contribution of the box diagrams and to a lesser
extent due to 
the energy dependence of the effective couplings. The deviation of the 
EBA cross section from the NLO result at large values of $\sqrt{\hat s}$
is the result of large
Sudakov-like electroweak logarithms of the form $\ln(\hat s/M_V^2)$
($V=W,\,Z$)~\cite{cc}. It is much more pronounced 
for $u\bar u\to e^+e^-$ than for $d\bar d\to e^+e^-$ which is due to the
different Feynman diagrams which contribute to the two initial states. 
In addition to the box diagrams with $Z$ boson exchange, only the 
crossed (direct) $W$ box diagram
contributes for up-type (down-type) quarks in the initial state (see 
Appendix~\ref{app:box}). As a result, for $d\bar d\to e^+e^-$, 
the contribution from the box diagrams is
less significant and the deviation of the EBA cross section from the
NLO result is not as large. 

\begin{figure}
\phantom{x}
\vskip 16.cm
\includegraphics{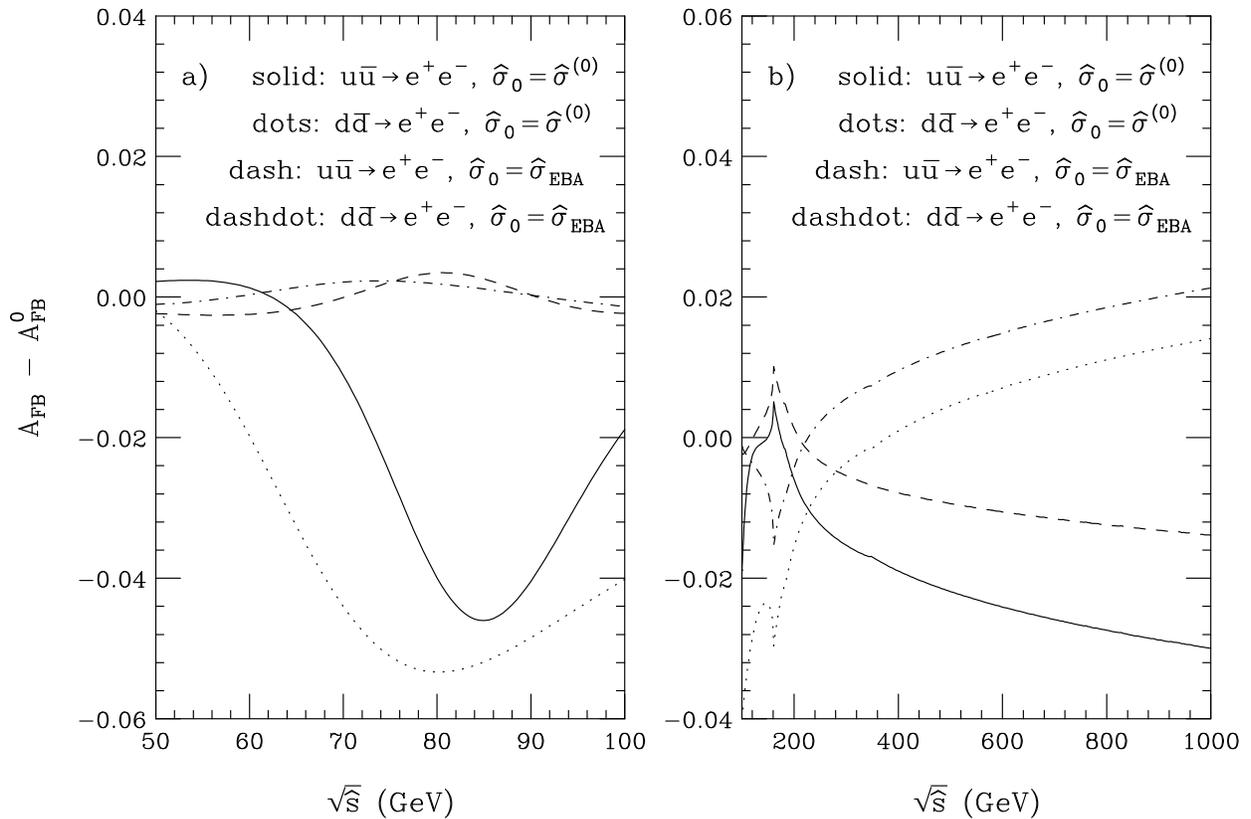}
\caption{The relative corrections to
the forward-backward asymmetry at parton level for $u\bar u \to e^+ e^-$ and 
$d \bar d \to e^+ e^-$ a) in the vicinity of the
$Z$ resonance and b) at high parton center of mass energies.}
\label{fig:z2}
\end{figure}

The forward-backward asymmetry at the parton level is given by 
\begin{equation}
A_{{\rm FB}} = \frac{\hat{\sigma}^F(\hat{s}) - \hat{\sigma}^B(\hat{s})}
              {\hat{\sigma}^F(\hat{s}) + \hat{\sigma}^B(\hat{s})}~,
\qquad 
\hat{\sigma}^{F\,(B)}(\hat{s}) = 
\int_{-\hat{s}/2\,(-\hat{s})}^{\,0\,(-\hat{s}/2)} d \hat t\; 
\frac{d\hat{\sigma}}{d \hat t}(\hat{s},\hat t)~,
\end{equation}
with $d\hat\sigma=d\hat\sigma^{(0)},
d\hat\sigma_{\rm{EBA}}, d \hat \sigma^{(0+1)}$ yielding the Born,
EBA and NLO predictions of $A_{{\rm FB}}$, respectively.  In
Fig.~\ref{fig:z2} we show the differences $A_{{\rm FB}}^{{\rm
NLO}}-A_{{\rm FB}}^{{\rm Born}}$ and $A_{{\rm FB}}^{{\rm NLO}}-A_{{\rm
FB}}^{{\rm EBA}}$ for the parton level processes $u\bar{u} \to
e^+e^-$ and $d\bar{d} \to e^+e^-$ in the vicinity of the
$Z$-peak (Fig.~\ref{fig:z2}a) and at large parton center of mass energies
(Fig.~\ref{fig:z2}b).  In the $Z$ peak region, $A_{{\rm FB}}$ 
is strongly reduced by the weak corrections. The reduction is more 
pronounced for the $d\bar d$ subprocess.  In the vicinity of the $Z$
boson peak, the EBA again provides a very good approximation. 
Due to the $\hat s$-dependence of the
effective couplings and the contributions from the box diagrams, 
the difference between $A_{{\rm FB}}^{{\rm
NLO}}$ and $A_{{\rm FB}}^{{\rm EBA}}$ rapidly increases in magnitude for
$\sqrt{\hat s}>100$~GeV.
At very high values of $\sqrt{\hat s}$, the weak corrections to
$u\bar u\to e^+e^-$ ($d \bar d\to e^+e^-$) considerably diminish 
(enhance) the forward-backward asymmetry.

\begin{figure}
\phantom{x}
\vskip 15.cm
\includegraphics{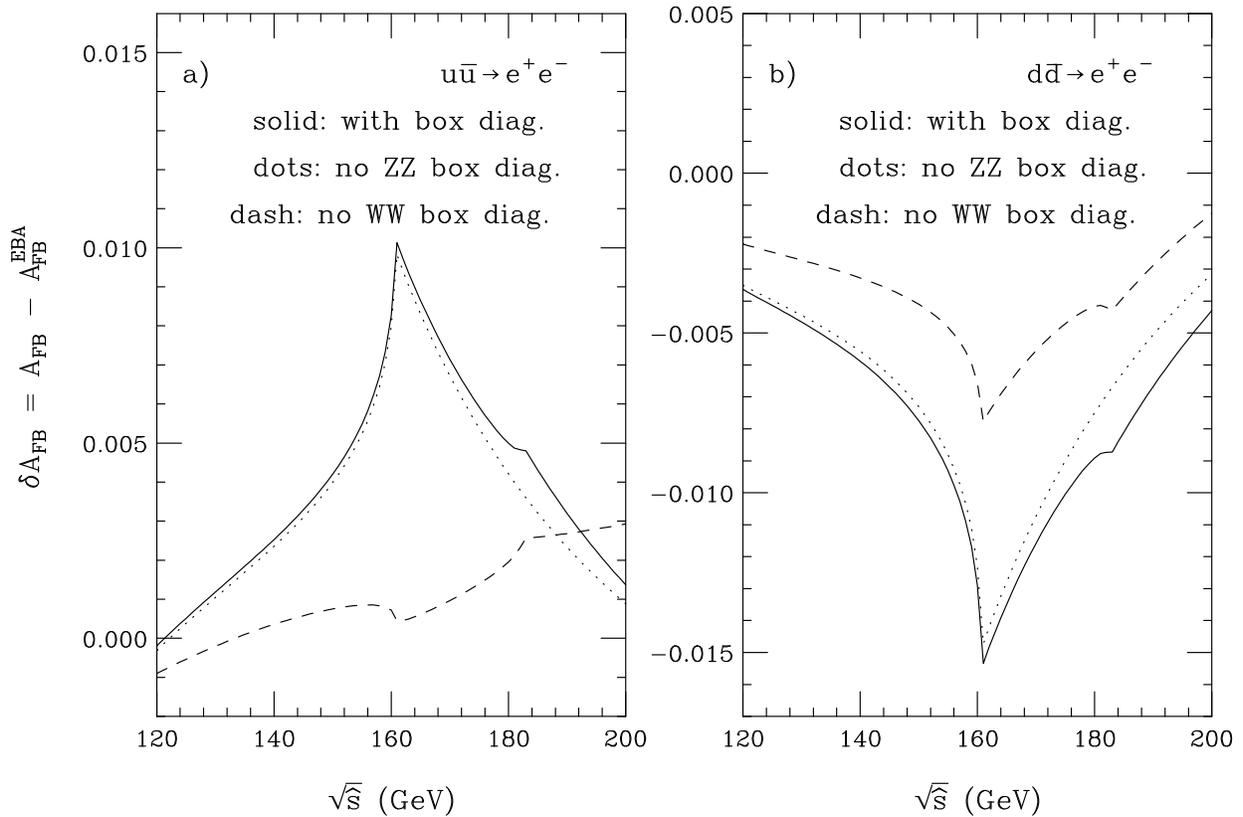}
\caption{The relative corrections to the forward-backward asymmetry 
at the parton level in the region around the $W$-pair production 
threshold, $\sqrt{\hat s}=2 M_W$, for a) $u \bar u \to e^+ e^-$  
and b) $d\bar d \to e^+ e^-$. The solid line shows $A_{\rm FB}-A_{\rm
FB}^{\rm EBA}$ when the full set of Feynman diagrams contributing to the
non-photonic weak corrections is taken into account. The dashed 
(dotted) lines show $A_{\rm FB}-A_{\rm FB}^{\rm EBA}$ when the $W$ ($Z$)
box diagrams are disregarded.}
\label{fig:z3}
\end{figure}

As can been seen from Fig.~\ref{fig:z2}b, the weak corrections also 
lead to a sharp peak (dip) in $A_{{\rm FB}}$ for $u\bar u\to e^+e^-$ 
($d \bar d\to e^+e^-$) in the vicinity of the $W^+W^-$ threshold, 
$\sqrt{\hat s}\approx 160$~GeV. The peak/dip is 
due to the $V-A$ nature of the coupling of the $W$ to fermions and 
a threshold effect in the $WW$ box and vertex diagrams. 
Figure~\ref{fig:z3} shows $\delta A_{\rm
FB}= A_{{\rm FB}}^{{\rm NLO}} - A_{{\rm FB}}^{{\rm EBA}}$ in more detail
in the region around the $W^+W^-$ threshold. The solid lines show
$\delta A_{\rm FB}$ including the full set of Feynman diagrams
contributing to the non-photonic weak corrections. Disregarding the $WW$
box diagrams removes a large portion of the peak/dip (dashed lines). The
remaining effect is due to initial and final state vertex corrections
involving two virtual $W$ bosons. For $u\bar u\to e^+e^-$ these
interfere destructively whereas for $d \bar d\to e^+e^-$ there is
constructive interference. The effect of the initial and final state
vertex corrections involving two virtual $W$ bosons therefore is more
pronounced for $d$-type quarks in the initial state.
Unlike the $W$ box diagrams, the $Z$ box graphs have
only a very small effect on the forward-backward asymmetry. This is due to the
small vector coupling of the $Z$ boson to the charged leptons. The
effect of removing the $Z$ box diagrams from the set of Feynman diagrams
describing the non-photonic weak corrections is shown by the dotted
lines in Fig.~~\ref{fig:z3}. 

%
%%% Section III: Numerical discussion %%%
%
\section{Phenomenological results}
\label{sec:numerics}

\subsection{Preliminaries}
\label{sec:prelim}

We shall now discuss the phenomenological implications of the ${\cal
O}(\alpha)$ genuine weak corrections to di-lepton production at the
Tevatron ($p\bar p$ collisions at $\sqrt{s}=2$~TeV) and the LHC ($pp$ 
collisions at $\sqrt{s}=14$~TeV). We first discuss the impact of the
non-universal weak corrections on the lepton pair invariant mass distribution
and the total cross section in the $Z$ pole region. We then consider how
the forward-backward asymmetry, $A_{\rm FB}$, is affected by these
corrections. The universal weak corrections are taken into account in
form of the effective Born approximation described in detail in
Sec.~\ref{sec:nonqed}. The SM parameters used in our numerical
simulations are listed in Eqs.~(\ref{eq:pars}) --~(\ref{eq:mhiggs}). 
To compute the cross section we use the MRSR2 set of parton distribution
functions~\cite{mrs}, and take the renormalization scale, $\mu$, and the
QED and QCD factorization scales, $\mu_{\rm QED}$ and $\mu_{\rm QCD}$, to be
$\mu^2=\mu_{\rm QED}^2=\mu_{\rm QCD}^2=\hat s$. 

To simulate detector acceptance, we impose the following transverse
momentum ($p_T$) and pseudo-rapidity ($\eta$) cuts ($l=e,\,\mu$)
\begin{equation}
p_T(l)>20~{\rm GeV,} \qquad |\eta(l)|<2.5. 
\label{eq:cuts}
\end{equation}
These cuts approximately model
the acceptance of the CDF II~\cite{cdfii} and D\O~\cite{dzero} detectors
at the Tevatron, and the ATLAS~\cite{atlas} and CMS~\cite{cms} detectors
at the LHC. Uncertainties in the energy measurements of the charged leptons 
in the detector are simulated in the calculation by Gaussian smearing 
of the particle four-momentum vector with standard deviation $\sigma$
which depends on the particle type and the detector. The numerical results 
presented here were calculated using $\sigma$ values based on the
CDF II and ATLAS specifications. 

The granularity of the detectors and the size of the electromagnetic
showers in the calorimeter make it difficult to discriminate between 
electrons and photons with a small opening angle. In such cases we
recombine the four-momentum vectors of the electron 
and photon to an effective electron four-momentum vector. The exact
recombination procedure is detector dependent. For calculations
performed at Tevatron energies we recombine the four-momentum vectors of
the electron and photon to an effective electron four-momentum vector if both
traverse the same calorimeter cell, assuming a calorimeter segmentation of
$\Delta\eta\times\Delta\phi=0.1\times 15^\circ$ ($\phi$ is the
azimuthal angle in the transverse plane). This procedure is similar to
that used by the CDF Collaboration in Run~I. The segmentation chosen
corresponds to that of the central part of the Run~I CDF calorimeter. At
LHC energies, we recombine the electron and photon four-momentum vectors
if their separation in the pseudorapidity -- azimuthal angle
plane, 
\begin{equation}
\Delta R(e,\gamma)=\sqrt{(\Delta\eta(e,\gamma))^2+(\Delta\phi(e,
\gamma))^2}, 
\end{equation}
is 
\begin{equation}
\Delta R(e,\gamma)< R_c=0.07, 
\end{equation}
similar to the resolution
expected for ATLAS~\cite{atlas}. Recombining the electron
and photon four-momentum vectors for small opening angles of the two
particles greatly reduces the effect of the mass singular logarithmic
terms associated with final state photon radiation~\cite{Baur:1998wa}. 

Muons are identified in a hadron collider detector by hits in the muon
chambers. In addition, one requires that the associated track is
consistent with a minimum ionizing particle. This limits the energy of a
photon to be smaller than a critical value $E^\gamma_c$ for small muon
-- photon opening angles. At the Tevatron we impose a 
$E_\gamma<E^\gamma_c=2$~GeV cut for photons traversing the same calorimeter
cell as the muon. At the LHC, following Ref.~\cite{atlas}, we require 
the photon energy to be smaller than $E^\gamma_c=5$~GeV if 
$\Delta R(\mu,\gamma)<0.3$. The cut on the photon energy increases the
size of the QED corrections for $m(\mu^+\mu^-)>100$~GeV~\cite{Baur:1998wa}.

We impose the cuts and lepton identification requirements described
above in all subsequent numerical simulations, unless explicitly 
noted otherwise. 

\subsection{Weak corrections to the di-lepton invariant mass
distribution and the $Z$ boson cross section}

QED corrections are known to have a profound impact on the shape of the 
di-lepton invariant mass distribution~\cite{Baur:1998wa}. Due to the
mass singular terms associated with final state photon radiation, the
differential cross section is reduced  in the $Z$ peak
region by about 10\% for electrons, and
by about 20\% for muons in the final state. Below the $Z$ resonance
region, final state photon radiation enhances the cross section by up to
a factor~1.5 with the maximum effect occurring at $m(l^+l^-)\approx
75$~GeV. For $m(l^+l^-)>100$~GeV, QED corrections reduce the $e^+e^-$
($\mu^+\mu^-$) differential cross section by about 5\% ($12-15\%$). In
contrast to final state photon radiation which significantly changes the
shape of the di-lepton invariant mass distribution, initial state QED
corrections are uniform and small ($\approx +0.4\%$). The distortion of
the Breit-Wigner shape of the $Z$ resonance curve due to final state QED
corrections causes the $Z$ boson mass extracted from data to be shifted
by about $-100$~MeV for electrons, and $-300$~MeV for muons, in the
final state~\cite{cdfwmass,d0wmass}. 

\begin{figure}
\phantom{x}
\vskip 14.cm
\includegraphics{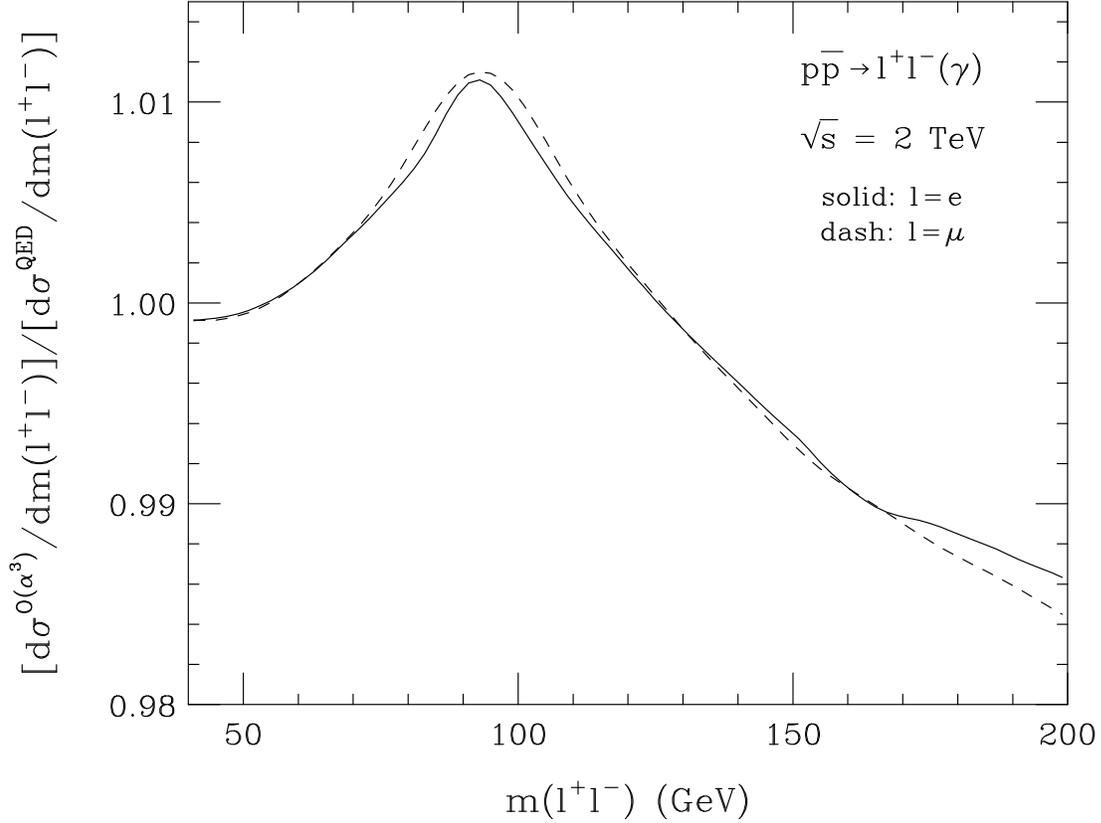}
\caption{The ratio $[d\sigma^{{\cal
O}(\alpha^3)}/dm(l^+l^-)]/[d\sigma^{\rm QED}/dm(l^+l^-)]$ as a function of
the di-lepton invariant mass at the Tevatron in the $Z$ peak region.
$\sigma^{{\cal O}(\alpha^3)}$ denotes the full NLO cross section, and 
$\sigma^{\rm QED}$ represents the cross section which includes the
factorizable electroweak corrections in form of the effective Born
approximation together with the ${\cal O}(\alpha)$ QED corrections. The
cuts and lepton identification requirements imposed are described in
Sec.~\ref{sec:prelim}. 
}
\label{fig:six}
\end{figure}

Based on the results obtained at the parton level (see
Sec.~\ref{sec:parton}), one expects that the non-universal ${\cal
O}(\alpha)$ weak corrections are small in the vicinity of the $Z$
resonance. Figure~\ref{fig:six} shows the ratio $[d\sigma^{{\cal
O}(\alpha^3)}/dm(l^+l^-)]/[d\sigma^{\rm QED}/dm(l^+l^-)]$ at the
Tevatron, where
$\sigma^{{\cal O}(\alpha^3)}$ denotes the full NLO cross section, and 
$\sigma^{\rm QED}$ represents the cross section which includes the
factorizable electroweak corrections in form of the effective Born
approximation together with the ${\cal O}(\alpha)$ QED corrections. Very
similar results are obtained at the LHC. For $m(l^+l^-)<50$~GeV, the
non-universal corrections are very small and negative. In the $Z$ peak
region, they enhance the differential cross section by up
1.2\%. Finally, for $m(l^+l^-)>130$~GeV, the non-universal weak
corrections become negative
and rapidly increase in magnitude. The small differences between the
results for electrons and muons in the final state are mostly due to the
different lepton identification requirements. The slight dip visible at
$m(l^+l^-)\approx 160$~GeV is caused by the $W$ pair threshold effect
discussed in Sec.~\ref{sec:parton}. 

Since the non-universal weak corrections are not uniform in the $Z$ peak
region, they 
are expected to shift the $Z$ boson mass extracted from data upward by 
several MeV. This is much smaller than the effect caused by QED
corrections but may not be negligible in future hadron collider
experiments. In order to make a more quantitative prediction of the
shift in $M_Z$ due to the weak corrections, detailed simulations which 
fully take into account detector response need to be performed. These are
beyond the scope of this paper.

The ratio, $R$, of the $W\to 
l\nu$ and $Z\to l^+l^-$ cross sections can be used to extract
information on the width of the $W$ boson~\cite{D0Wcross,cdfr}. Since 
the QCD corrections to $W$ and $Z$ production
are very similar, they cancel almost perfectly in the $W$ to $Z$ cross
section ratio; the ${\cal O}(\alpha_s)$ corrections to $R$ are of
${\cal O}(1\%)$ or less, depending on the set of parton distribution
functions used~\cite{willy}. In addition many experimental uncertainties
such as the luminosity uncertainty, cancel in the cross section ratio. 
Accurate knowledge of how electroweak corrections affect the $W\to l\nu$
and the di-lepton cross sections in the $Z$ resonance region is thus
very important. 

The effect of the non-universal weak corrections on the $l^+l^-$
invariant mass distribution is also reflected in the total cross section
in the $Z$ resonance region. 
In Table~\ref{tab:one} we list the cross section ratios 
\begin{equation}
K^{EW}={\sigma^{{\cal O}(\alpha^3)}\over\sigma^{\rm EBA}}~,
\end{equation}
(``EW $K$-factor'') and 
\begin{equation}
K^{\rm QED}={\sigma^{\rm QED}\over\sigma^{\rm EBA}}
\end{equation}
(``QED $K$-factor'') for $75~{\rm GeV}<m(l^+l^-)<105$~GeV ($l=e,\,\mu$)
at the Tevatron 
with and without taking the cuts and lepton identification requirements
described in Sec.~\ref{sec:prelim} into account. Similar results are
obtained at the LHC. One observes that the
genuine weak interactions increase the cross section by about
1.0\%. Approximately one-half of the enhancement is due to the ${\cal
O}(g^4m_t^2/M_W^2)$ corrections to $\sin^2\theta_{\rm eff}^l$ and
$M_W$. In
contrast, the QED corrections decrease the cross section. The size of
the QED corrections to the cross section depends on the flavor of the
final state lepton and whether cuts and lepton identification
requirements are taken into account or not. Without detector effects
taken into account, the QED corrections are numerically more important
than the genuine weak corrections. Due to the mass singular terms
associated with final state photon radiation, the QED corrections for
the $e^+e^-$ final state are larger than those in the muon case. The
full electroweak corrections reduce the $e^+e^-(\gamma)$
($\mu^+\mu^-(\gamma)$)  cross section in the $Z$ resonance region by
about 5\% (2\%). 

The recombination of electron and photon momenta when the
opening angle between the two particles is small strongly reduces the 
effect of the QED corrections to the integrated $e^+e^-$ cross
section. As a result, the effects of the QED corrections and the genuine weak
corrections partially cancel. The net effect of the electroweak
corrections is a decrease of the cross section by 1.2\%. In the muon 
case, lepton identification requirements increase the magnitude of the 
QED corrections, and the full electroweak corrections decrease the cross
section by more than 6\%. 

\begin{figure}
\phantom{x}
\vskip 14.cm
\includegraphics{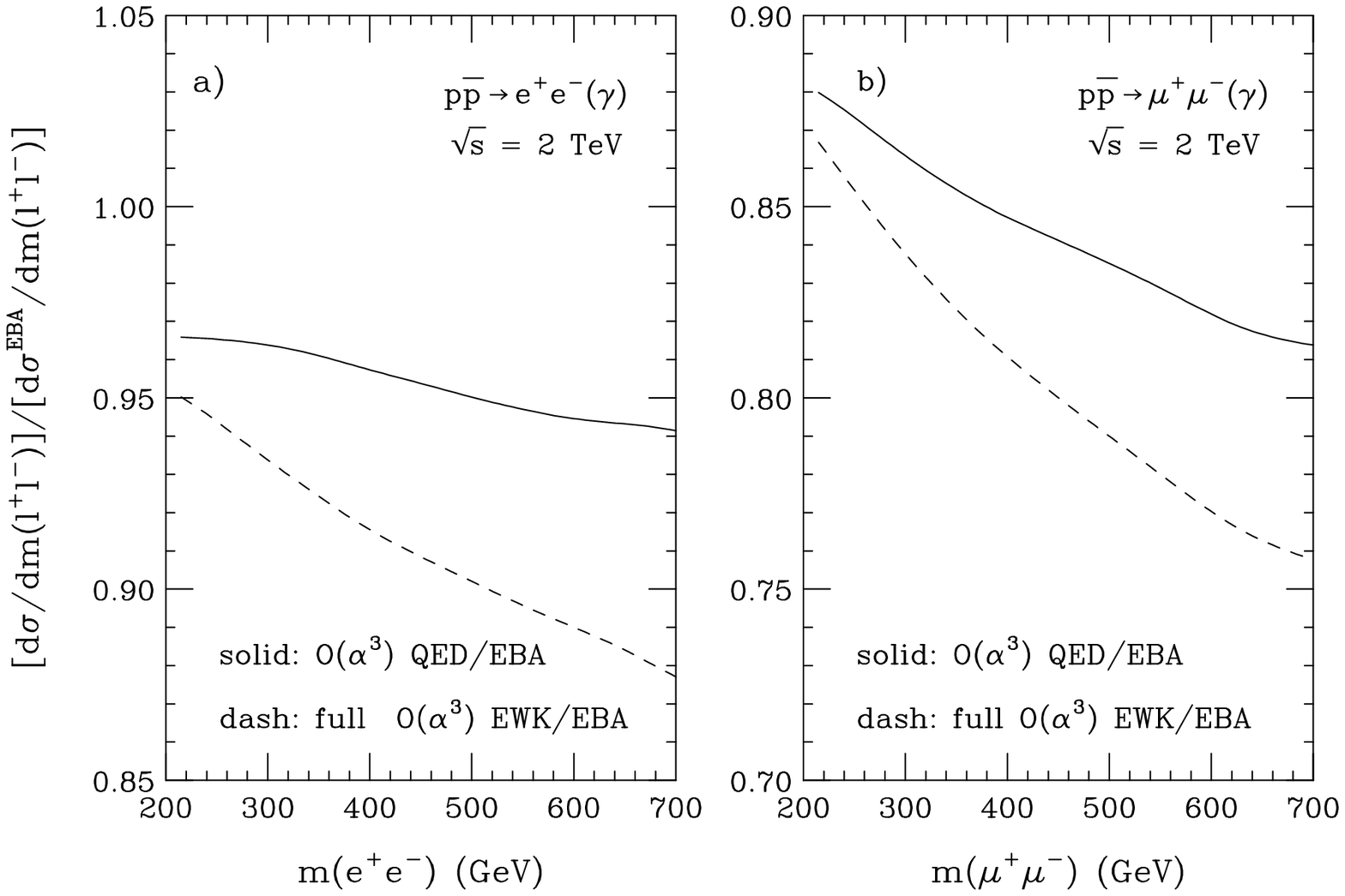}
\caption{The ratio $[d\sigma/dm(l^+l^-)]/[d\sigma^{\rm EBA}/dm(l^+l^-)]$ 
as a function of the di-lepton invariant mass for a) $p\bar p\to
e^+e^-(\gamma)$ and b) $p\bar p\to\mu^+\mu^-(\gamma)$ at
$\sqrt{s}=2$~TeV. The dashed lines show the ratio of
the complete ${\cal O}(\alpha^3)$ electroweak and the EBA differential
cross section. The solid lines display the corresponding ratio for the 
case where only the ${\cal O}(\alpha)$ QED corrections and the 
factorizable electroweak corrections in form of the effective Born
approximation are taken into account. The
cuts and lepton identification requirements imposed are described in
Sec.~\ref{sec:prelim}. 
}
\label{fig:fig7}
\end{figure}

\begin{figure}
\phantom{x}
\vskip 14.cm
\includegraphics{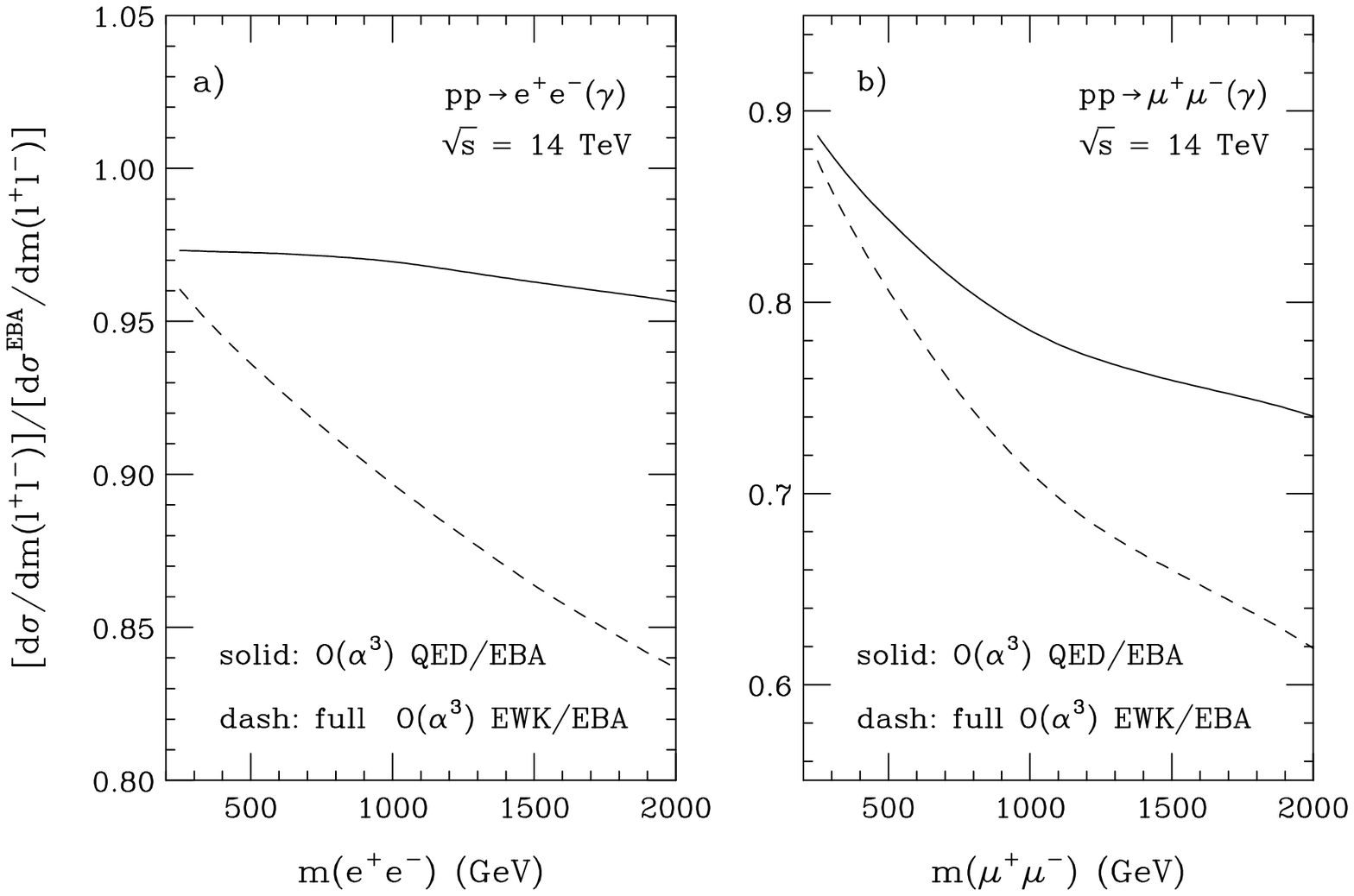}
\caption{The ratio $[d\sigma/dm(l^+l^-)]/[d\sigma^{\rm EBA}/dm(l^+l^-)]$ 
as a function of the di-lepton invariant mass for a) $pp\to
e^+e^-(\gamma)$ and b) $pp\to\mu^+\mu^-(\gamma)$ at
$\sqrt{s}=14$~TeV. The dashed lines show the ratio of
the complete ${\cal O}(\alpha^3)$ electroweak and the EBA differential
cross section. The solid lines display the corresponding ratio for the 
case where only the ${\cal O}(\alpha)$ QED corrections and the 
factorizable electroweak corrections in form of the effective Born
approximation are taken into account. The
cuts and lepton identification requirements imposed are described in
Sec.~\ref{sec:prelim}. 
}
\label{fig:fig8}
\end{figure}

The effect of the non-factorizable weak corrections on the di-lepton
invariant mass distribution at the Tevatron for large values of 
$m(l^+l^-)$ is shown in Fig.~\ref{fig:fig7} where we plot the ratio of
the complete ${\cal O}(\alpha^3)$ electroweak and the EBA differential
cross section as a function of $m(l^+l^-)$ (dashed lines). In order to 
make the effect of the non-factorizable weak corrections more
transparent, we also show the corresponding ratio for the case where
only the ${\cal O}(\alpha)$ QED corrections and the 
factorizable electroweak corrections in form of the effective Born
approximation are taken into account (solid
lines). Figure~\ref{fig:fig8} shows the corresponding results for the
LHC. Due to the recombination of electrons and photons, the QED
corrections reduce the $e^+e^-$ differential cross section by only $3-5\%$
over the invariant mass regions considered. In the muon case, the cut on
the photon energy for photons which have a small opening angle with the
muon reduces the hard photon part of the ${\cal O}(\alpha^3)$
$\mu^+\mu^-(\gamma)$ cross section. As a result, the QED corrections are
much more pronounced and display a much stronger dependence on the
di-lepton invariant mass than in the $e^+e^-$ case. The non-factorizable
weak corrections are seen to increase rapidly in size with $m(l^+l^-)$. As
mentioned before (see Sec.~\ref{sec:parton}), this is due to the
presence of Sudakov-like electroweak logarithms of the form
$\ln(m(l^+l^-)/M_V)$ ($V=W,\,Z$). Most of the effect is
caused by the contribution of up-type quarks in the initial state (see
Fig.~\ref{fig:z1}). For $m(e^+e^-)=500$~GeV ($m(e^+e^-)=1.5$~TeV), the
electroweak corrections reduce the cross section by about 10\% (15\%) at
the Tevatron (LHC). In the muon channel, the ${\cal O}(\alpha)$ 
electroweak corrections are larger in magnitude than the ${\cal
O}(\alpha_s)$ QCD corrections for $\mu^+\mu^-$ invariant masses larger
than about 500~GeV. For $m(\mu^+\mu^-)=2$~TeV at the LHC, the ${\cal
O}(\alpha)$ electroweak radiative corrections  reduce the cross
section by more than 35\%, which is approximately equal to the expected 
statistical uncertainty in a 200~GeV bin centered at 
$m(\mu^+\mu^-)=2$~TeV for 100~fb$^{-1}$. It will thus be important to
take into account the non-factorizable weak corrections when measuring
the Drell-Yan cross section at large di-lepton invariant masses at the
LHC. 

The results shown in Fig.~\ref{fig:fig8} should be interpreted with
caution. Since the non-factorizable weak corrections become large for 
di-lepton invariant masses above 1~TeV, they need to be resummed in order to
obtain accurate predictions in this phase space region (for a
recent review of the resummation of electroweak Sudakov-like
logarithms see Ref.~\cite{melles}). A calculation of Drell-Yan 
production in hadronic collisions which includes resummation of electroweak
logarithms has not been carried out yet. 

\subsection{Weak corrections to the forward-backward asymmetry}
\label{sec:afb}

We now discuss how the non-universal weak corrections affect the
forward-backward asymmetry, $A_{\rm FB}$. The expressions used in the
literature to define $A_{\rm FB}$ at the Tevatron~\cite{CDFAFB} and the
LHC~\cite{Baur:1998wa} are slightly different. 
For $p\bar p$ collisions at Tevatron energies, $A_{\rm FB}$ usually is 
defined by 
\begin{equation}
A_{\rm FB}={F-B\over F+B}~,
\label{EQ:DEFAFB}
\end{equation}
where
\begin{equation}
F=\int_0^1{d\sigma\over d\cos\theta^*}\,d\cos\theta^*, \qquad
B=\int_{-1}^0{d\sigma\over d\cos\theta^*}\,d\cos\theta^*.
\label{EQ:DEFFB}
\end{equation}
Here, $\cos\theta^*$ is given by~\cite{CDFAFB,CS}
\begin{equation}
\cos\theta^*={2\over m(l^+l^-)\sqrt{m^2(l^+l^-)+p_T^2(l^+l^-)}}\left 
[p^+(l^-)p^-(l^+)-p^-(l^-)p^+(l^+)\right ]
\label{EQ:CSTAR}
\end{equation}
with
\begin{equation}
p^\pm={1\over\sqrt{2}}\left (E\pm p_z\right ),
\end{equation}
where $E$ is the energy and $p_z$ is the longitudinal component of the
momentum vector. In this definition of $\cos\theta^*$, the polar axis 
is taken to be the bisector of the
proton beam momentum and the negative of the anti-proton beam momentum
when they are boosted into the $l^+l^-$ rest frame. In $p\bar p$ 
collisions at Tevatron energies, the flight direction of the incoming 
quark coincides with the proton beam direction for a large fraction of
the events. The definition of $\cos\theta^*$ in Eq.~(\ref{EQ:CSTAR})
has the advantage of minimizing the effects of the QCD corrections (see
below). In the limit of vanishing di-lepton $p_T$,
$\theta^*$ coincides with the angle between the lepton and the incoming 
proton in the $l^+l^-$ rest frame. 

\begin{figure}
\phantom{x}
\vskip 14.cm
\includegraphics{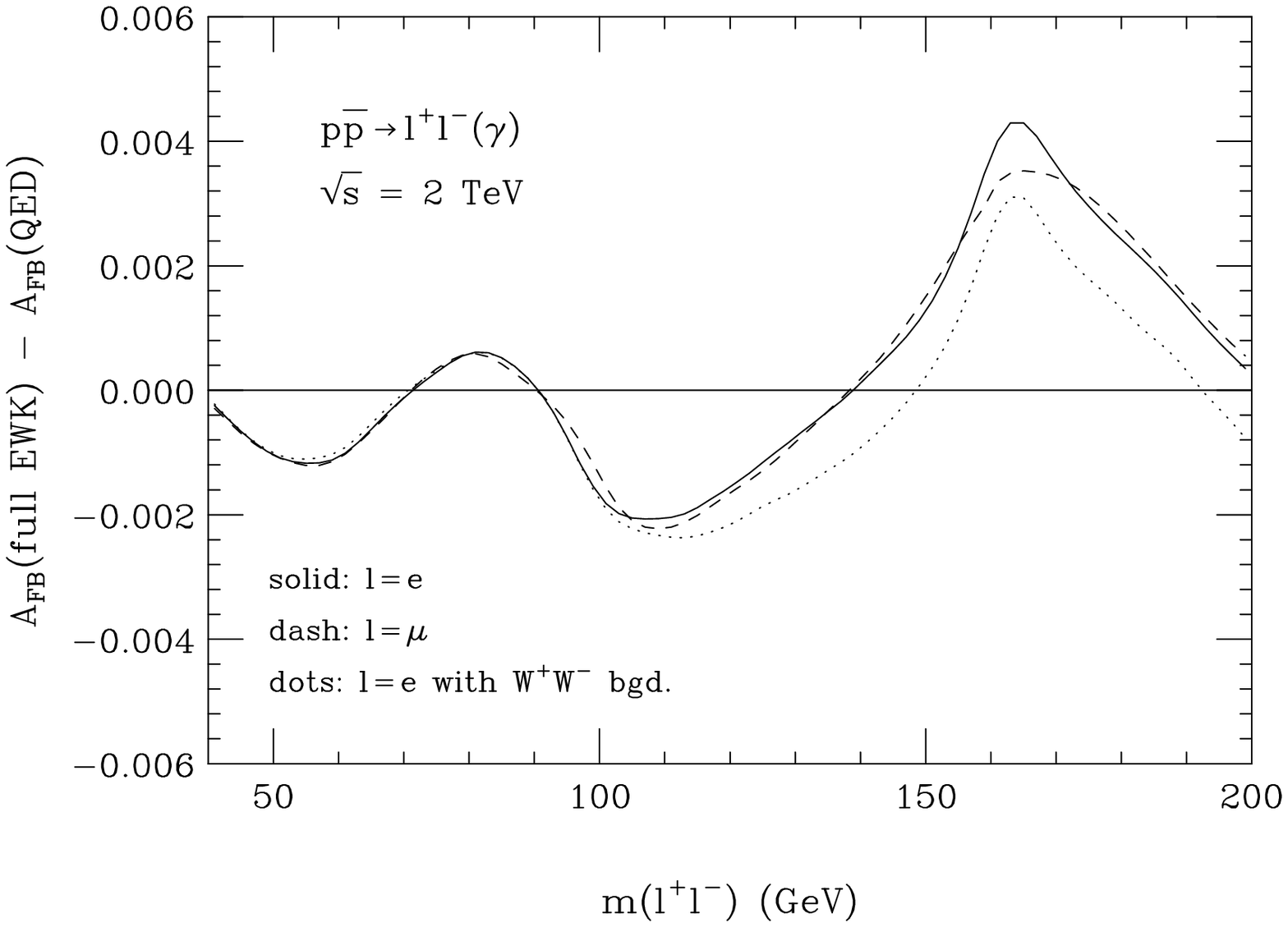}
\caption{The difference $A_{\rm FB}({\rm full~EWK})-A_{\rm FB}({\rm QED})$ for
$p\bar p\to l^+l^-(\gamma)$ at $\sqrt{s}=2$~TeV. $A_{\rm FB}({\rm full~EWK})$
denotes the forward-backward asymmetry calculated 
taking the full ${\cal O}(\alpha)$
electroweak corrections and the ${\cal O}(g^4m_t^2/M_W^2)$ corrections
into account. $A_{\rm FB}({\rm QED})$ only includes the
${\cal O}(\alpha)$ QED corrections, in addition to the factorizable
corrections absorbed in the EBA. The solid and dashed lines show the
results for electron and muon final states, respectively. The dotted
line shows the difference in the asymmetry taking the $p\bar p\to
W^+W^-\to e^+e^-p\llap/_T$ background with $p\llap/_T<20$~GeV in
$A_{\rm FB}({\rm full~EWK})$ into
account. Additional cuts and the lepton identification requirements imposed 
are described in Sec.~\ref{sec:prelim}. 
}
\label{fig:fig9}
\end{figure}

QED corrections are known to have a significant effect on the forward
backward asymmetry for $50~{\rm GeV}<m(l^+l^-)<90$~GeV but are small for
di-lepton masses larger than 100~GeV~\cite{Baur:1998wa}. The difference
\begin{equation}
\Delta A_{\rm FB}=A_{\rm FB}({\rm full~EWK})-A_{\rm FB}({\rm QED}) 
\end{equation}
is the quantity which
best displays how the non-universal weak interactions influence the
forward-backward asymmetry. Here,
$A_{\rm FB}({\rm full~EWK})$ is the forward-backward asymmetry calculated 
taking the full ${\cal O}(\alpha)$
electroweak corrections and the ${\cal O}(g^4m_t^2/M_W^2)$ corrections
into account. $A_{\rm FB}({\rm QED})$, on the other hand, only includes the
${\cal O}(\alpha)$ QED corrections, in addition to the factorizable
corrections absorbed in the EBA. Fig.~\ref{fig:fig9} shows $\Delta A_{\rm FB}$ 
for di-lepton masses between 40~GeV and 200~GeV at the Tevatron.  
It demonstrates that
the weak corrections have only a small effect on the forward-backward
asymmetry for $m(l^+l^-)<200$~GeV. The peak in $\Delta A_{\rm FB}$
located at $m(l^+l^-)\approx 
160$~GeV originates from threshold effects associated with the $W$ box
diagrams (see Figs.~\ref{fig:z2}b and~\ref{fig:z3}). Although the
contributions from up- and down-type quarks in the initial state tend to
cancel, a significant effect remains since the $u\bar u$ parton luminosity is
much larger than that for $d \bar d$ pairs for the di-lepton invariant
mass range of interest. The slight differences between electron and muon
final states in $\Delta A_{\rm FB}$ originate from the different detector
resolutions for the two final states. The peak in $\Delta A_{\rm FB}$
located at $m(l^+l^-)\approx 80$~GeV is also associated with the $WW$ box
diagrams. In this region one of the $W$ bosons in the loop is on-shell,
causing a small resonance like enhancement in the forward-backward
asymmetry. 

The peak located at $m(l^+l^-)\approx 160$~GeV is a characteristic 
signature of the non-factorizable weak interactions in Drell-Yan 
production and it is interesting to investigate whether it may be 
observable in Run~II. Since the size of the effect is small, one has to
worry about how higher QCD corrections, detector effects and background
processes affect the peak. 
QCD corrections are uniform in the region of interest and do not
modify the structure of the peak~\cite{kamal}. 
Detector resolution effects broaden the peak and reduce its
height. These effects are taken into account in
Fig.~\ref{fig:fig9}. Higher order Coulomb corrections are also expected
to modify the shape of the peak. Finally, backgrounds from $t\bar
t\to l^+l^-\nu\bar\nu b\bar b$, $W^+W^-\to l^+l^-\nu\bar\nu$ and $ZZ\to
l^+l^-\nu\bar\nu$ have to be taken into account. The $\bar tt$ and 
$W^+W^-$ backgrounds can either be
subtracted using the experimentally determined $e\mu p\llap/_T+X$ cross
section, or suppressed by imposing missing transverse momentum and jet
veto cuts. Requiring that no jets with $p_T(j)>20$~GeV and
$|\eta(j)|<3.5$ are observed and imposing a $p\llap/_T<20$~GeV cut
suppresses the $ZZ$ and $\bar tt$ backgrounds to negligible levels at
Tevatron energies. $\Delta A_{\rm FB}$ for the $e^+e^-$ final state
including the 
contribution of the $W^+W^-$ background is shown by the dotted line in
Fig.~\ref{fig:fig9}. The $W^+W^-$ background is completely negligible
for $m(e^+e^-)<100$~GeV. For larger invariant masses it 
slightly decreases $\Delta A_{\rm FB}$ but does not change the shape of the 
peak at 160~GeV. Similar results are obtained for muons in the final
state. Detector effects and background processes thus will have little 
effect on the observability of the peak originating from the $WW$ box
diagrams. 

A simple method to estimate whether one can hope to observe the peak at
160~GeV in Run~II is to compare 
the statistical uncertainty expected for $A_{\rm FB}$ in
a 10~GeV bin centered at 160~GeV with the variation of the forward-backward
asymmetry due to the non-factorizable weak corrections in the same
region. For 20~fb$^{-1}$, the statistical uncertainty in the
$160\pm 5$~GeV bin is found to be
$\delta A_{\rm FB}(stat)\approx 0.016$ per lepton channel and experiment, 
whereas the non-factorizable weak
corrections change $A_{\rm FB}$ by about~0.003. It will thus be difficult to
observe the threshold effect in the $WW$ box diagrams in Run~II.

For the definition of $\cos\theta^*$ given in Eq.~(\ref{EQ:CSTAR}),
$A_{\rm FB}=0$ for $pp$ collisions. The easiest way to obtain a non-zero
forward-backward asymmetry at the LHC is to extract the 
quark direction in the initial state from the boost direction of the 
di-lepton system with respect 
to the beam axis~\cite{Dittmar}. The cosine of the angle between the 
lepton and the quark in the $l^+l^-$ rest frame is then
approximated by
\begin{equation}
\cos\theta^*={|p_z(l^+l^-)|\over p_z(l^+l^-)}~{2\over 
m(l^+l^-)\sqrt{m^2(l^+l^-)
+p_T^2(l^+l^-)}}\left [p^+(l^-)p^-(l^+)-p^-(l^-)p^+(l^+)\right ].
\label{EQ:CSTAR1}
\end{equation}

At the LHC, the sea -- sea quark flux is much larger than at the
Tevatron.
As a result, the probability, $f_q$, that the quark direction and the
boost direction of the di-lepton system coincide is significantly smaller 
than one. The forward-backward asymmetry is therefore smaller than at
the Tevatron.
Events with a large rapidity of the di-lepton system, $y(l^+l^-)$,
originate from collisions where at least one of the partons carries a
large fraction $x$ of the proton momentum. Since valence quarks
dominate at high values of $x$, a cut on the di-lepton rapidity 
increases $f_q$, and thus the asymmetry~\cite{Dittmar} and the
sensitivity to the effective weak mixing angle. In the following we
therefore impose a 
\begin{equation}
|y(l^+l^-)|>1
\label{EQ:RCUT}
\end{equation}
cut in all numerical calculations of the forward-backward asymmetry at
the LHC. 

For muons in the final state we impose the $p_T$ and pseudo-rapidity
cuts listed in Eq.~(\ref{eq:cuts}). In the $e^+e^-$ case, we allow one of
the electrons to be in the range $|\eta(e)|<4.9$, whereas the other
electron is required to be within $|\eta(e)|<2.5$. This takes into
account the possibility of using the forward calorimeter in ATLAS for electron
identification. The standard rapidity coverage of the ATLAS and
CMS detectors (see Eq.~(\ref{eq:cuts})) for leptons, $|\eta(l)|<2.5$, 
is known to significantly reduce
$A_{\rm FB}$~\cite{Baur:1998wa}. In addition, it results in a reduction of
the total cross section in the $Z$ pole region by roughly a
factor~5. Combined, these effects greatly reduce the chances for a
precise measurement of the weak mixing angle at the LHC. As demonstrated
in Ref.~\cite{Haywood:1999qg}, a large fraction of the sensitivity lost
can be recovered if one can make use of the forward calorimeter to detect
one of the electrons. 

\begin{figure}
\phantom{x}
\vskip 14.cm
\includegraphics{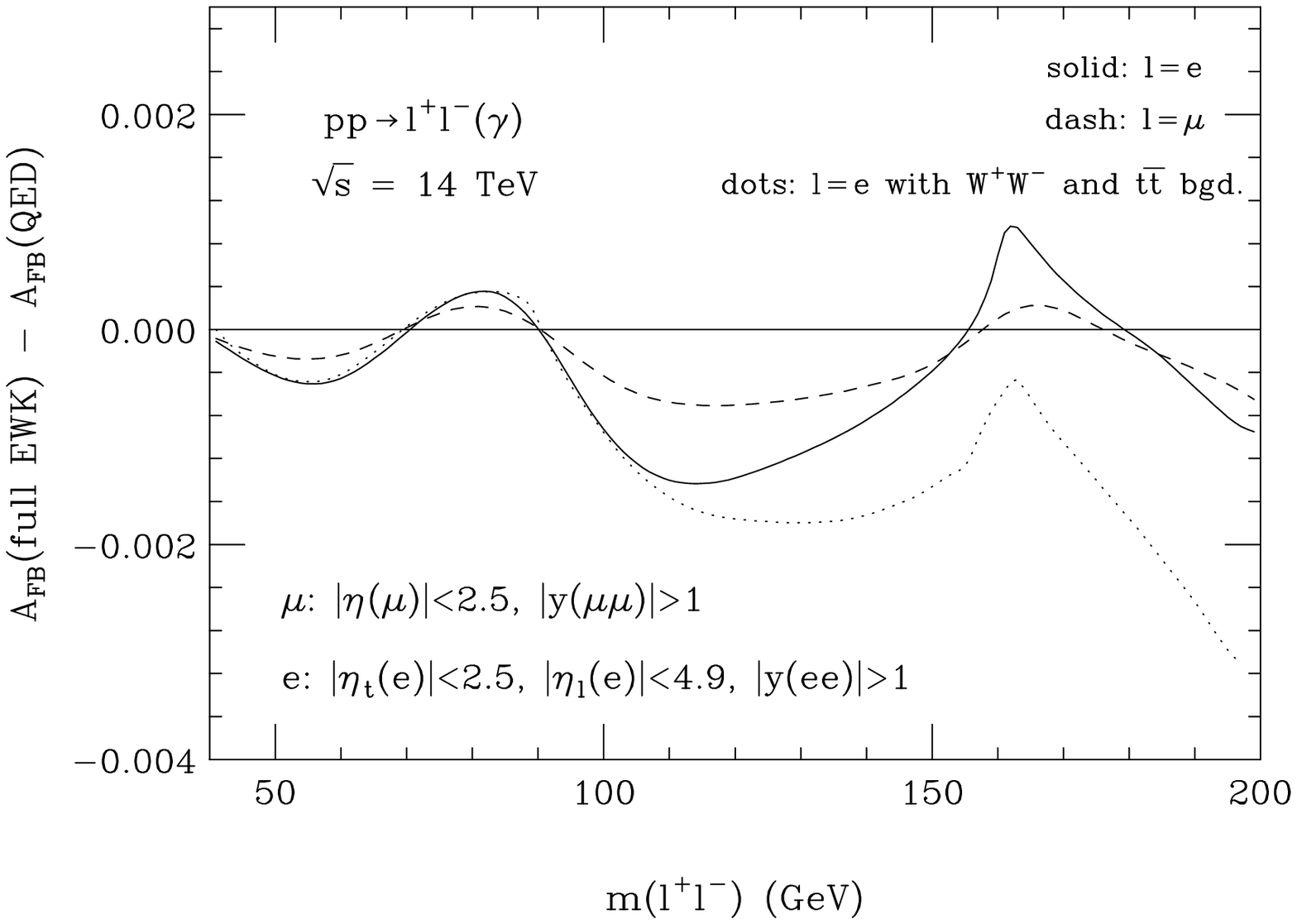}
\caption{The difference $A_{\rm FB}({\rm full~EWK})-A_{\rm FB}({\rm QED})$ for
$pp\to l^+l^-(\gamma)$ at $\sqrt{s}=14$~TeV. $A_{\rm FB}({\rm full~EWK})$
denotes the forward-backward asymmetry calculated 
taking the full ${\cal O}(\alpha)$
electroweak corrections and the ${\cal O}(g^4m_t^2/M_W^2)$ corrections
into account. $A_{\rm FB}({\rm QED})$ only includes the
${\cal O}(\alpha)$ QED corrections, in addition to the factorizable
corrections absorbed in the EBA. The solid and dashed lines show the
results for electron and muon final states, respectively. The dotted
line shows the difference in the asymmetry taking the $pp\to
W^+W^-\to e^+e^-p\llap/_T$ and $pp\to\bar tt\to e^+e^-p\llap/_T\bar bb$ 
background in $A_{\rm FB}({\rm full~EWK})$ into account. The cuts and lepton
identification requirements
imposed are described in Secs.~\ref{sec:prelim} and~\ref{sec:afb}. 
}
\label{fig:fig10}
\end{figure}

$\Delta A_{\rm FB}$ at the LHC for $40~{\rm GeV}<m(l^+l^-)<200$~GeV is shown
in Fig.~\ref{fig:fig10}. In the muon channel, $\Delta A_{\rm FB}$ is
smaller than 0.001 in magnitude over the entire mass range considered
and the peak caused by the threshold effects associated with the $WW$
box diagrams is significantly washed out (dashed line). For electrons,
on the other hand, the peak at $m(e^+e^-)\approx 160$~GeV is quite
pronounced and $\Delta A_{\rm FB}$ is roughly a factor~2 larger than in the
muon case. The dotted line shows the result for electrons, taking into
account electroweak background processes. To reduce the $\bar tt$,
$W^+W^-$ and $ZZ$ backgrounds we require that $p\llap/_T<20$~GeV, and
that no jets with $p_T(j)>50$~GeV and $|\eta(j)|<5$ are observed. As at
the Tevatron, the $ZZ$ background is negligible. However, since
the $\bar tt$ cross section at the LHC is more than a factor~100 larger
than at the Tevatron, the $\bar tt$ background is much more important at the
LHC and cannot be neglected. For $m(e^+e^-)>100$~GeV, the electroweak
background processes are 
seen to significantly modify $\Delta A_{\rm FB}$, however, without affecting
the shape of the peak. The expected statistical errors for the forward
backward asymmetry in a 10~GeV bin centered at $m(l^+l^-)=160$~GeV for
an integrated luminosity of 100~fb$^{-1}$ (1~ab$^{-1}$) are $\delta
A_{\rm FB}(stat)=0.0036$ ($\delta A_{\rm FB}(stat)=0.0011$) in the
electron, and 
$\delta A_{\rm FB}(stat)=0.0062$ ($\delta A_{\rm FB}(stat)=0.0020$) in
the muon channel. From Fig.~\ref{fig:fig10} it is then clear that
observation of the peak caused by threshold effects associated with the
$WW$ box diagrams at the LHC will also be quite difficult. 

In the $Z$ peak region, the forward-backward asymmetry, $A_{\rm FB}$, 
provides a tool to measure $\sin^2\theta_{\rm eff}^l$. In this region, 
the forward-backward asymmetry can to a very good 
approximation be parameterized by~\cite{Rosner}
\begin{equation}
A_{\rm FB}=b\left(a-\sin^2\theta^{l}_{\rm eff}\right),
\label{EQ:AFB}
\end{equation}
both in the Born approximation and including ${\cal O}(\alpha)$
electroweak corrections. The parameter $b$ controls the sensitivity of
$A_{\rm FB}$ to the effective weak mixing angle. The values of the 
coefficients $a$ and $b$ in the EBA, and the shifts introduced by the QED
and the non-universal weak 
corrections are listed in Table~\ref{tab:two} for the integrated forward
backward asymmetry in the region $75~{\rm GeV}< m(l^+l^-)<105$~GeV. The
coefficients $a$ and $b$ including the full ${\cal O}(\alpha)$
electroweak corrections are given by
\begin{equation}
a^{{\cal O}(\alpha^3)}=a^{\rm QED}+\Delta a^{\rm weak}, \qquad
b^{{\cal O}(\alpha^3)}=b^{\rm QED}+\Delta b^{\rm weak}.
\end{equation}
Here
\begin{equation}
a^{\rm QED}=a^{\rm EBA}+\Delta a^{\rm QED},\qquad
b^{\rm QED}=b^{\rm EBA}+\Delta b^{\rm QED},
\end{equation}
denote the parameters obtained when only the QED corrections and the
factorizable electroweak corrections in form of the effective Born
approximation are taken into account.

Since previous measurements of $A_{\rm FB}$ at the Tevatron~\cite{CDFAFB}
have corrected for detector effects, we do not impose any cuts
or lepton identification requirements when extracting $a$ and $b$ for
$p\bar p$ collisions at $\sqrt{s}=2$~TeV. For the LHC we
impose the cuts and lepton identification requirements described above and
in Sec.~\ref{sec:prelim}. In order to obtain the results listed in
Table~\ref{tab:two}, we have varied the Higgs boson mass between 75~GeV
and 350~GeV, corresponding to a variation of $\sin^2\theta^{l}_{\rm eff}$
between 0.23149 and 0.23225.

Table~\ref{tab:two} shows that the non-factorizable weak interactions
have only a small effect on $a$ and $b$. QED corrections, on the other
hand, have a significant impact. In particular they reduce the
sensitivity of $A_{\rm FB}$ to the effective weak mixing angle. The rather
large differences between the coefficients $a$ and $b$ for
$e^+e^-(\gamma)$ and $\mu^+\mu^-(\gamma)$ final states at the LHC are
due to the different rapidity coverage assumed for electrons and muons.
The sensitivity of $A_{\rm FB}$ to $\sin^2\theta^{l}_{\rm eff}$ at the
Tevatron is significantly higher than at the LHC~\cite{Baur:1998wa}. 

At the Tevatron in Run~II, one expects to measure $\sin^2\theta^{l}_{\rm eff}$
with a precision of 0.0005 (0.0006) in the electron (muon) channel,
assuming an integrated luminosity of 10~fb$^{-1}$~\cite{Brock:1999ep}. 
At the LHC, with 100~fb$^{-1}$ and the rapidity cuts described above,
one hopes to reach an accuracy of 0.00014 for the effective weak mixing
angle in the $e^+e^-(\gamma)$ final state~\cite{Haywood:1999qg}.  
Ignoring the ${\cal O}(\alpha)$ electroweak radiative corrections would
shift $\sin^2\theta^{l}_{\rm eff}$ by $(2-3)\, 10^{-4}$
($(3-5)\,10^{-4}$) towards smaller (larger) values at the Tevatron
(LHC). The shift is of the same size (Tevatron) or larger (LHC) than the
expected experimental uncertainty. It will thus 
be necessary to take ${\cal O}(\alpha)$ corrections into account when
one extracts the effective weak mixing angle from Drell-Yan production 
in future Tevatron or LHC experiments. If only the
non-universal weak corrections are neglected, the calculated and the true 
value of $\sin^2\theta^{l}_{\rm eff}$ deviate by $(2-3)\, 10^{-5}$ 
($(1-2)\, 10^{-5}$). Non-universal weak corrections thus have a small or
negligible effect on the effective weak mixing angle extracted from the
forward-backward asymmetry. 

\begin{figure}
\phantom{x}
\vskip 14.cm
\includegraphics{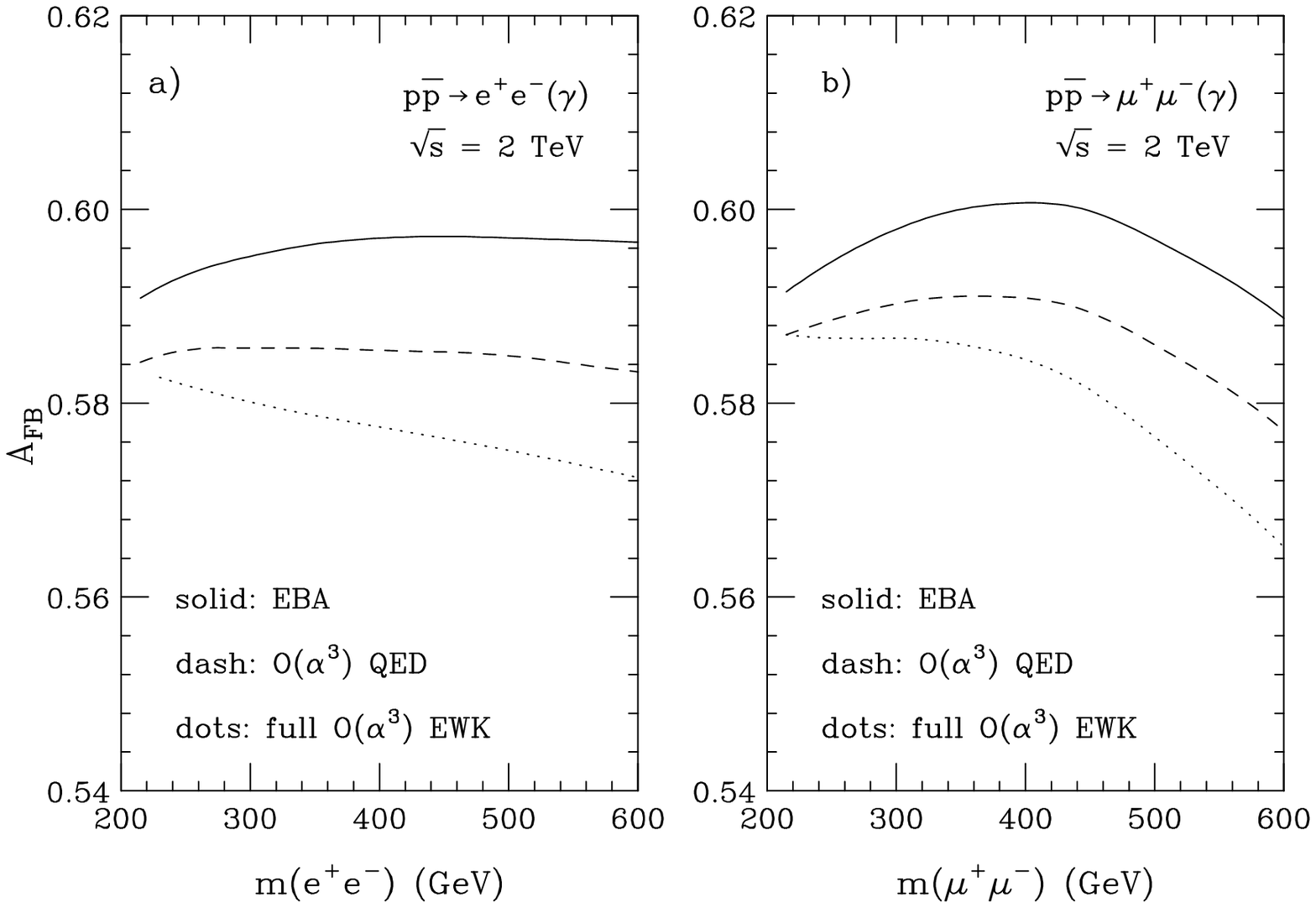}
\caption{The forward-backward asymmetry as a function of $m(l^+l^-)$ for
a) $p\bar p\to e^+e^-(\gamma)$ and b) $p\bar p\to \mu^+\mu^-(\gamma)$ at
$\sqrt{s}=2$~TeV. Shown are the asymmetry in the EBA (solid lines),
including pure QED corrections in addition to those corrections which
are part of the EBA (dashed lines), and the asymmetry 
taking the complete set of ${\cal O}(\alpha)$
electroweak corrections and the ${\cal O}(g^4m_t^2/M_W^2)$ corrections
into account (dotted lines). The cuts and lepton identification requirements
imposed are described in Sec.~\ref{sec:prelim}. 
}
\label{fig:fig11}
\end{figure}

Analogously to the $l^+l^-$ invariant mass distribution, one expects
that the genuine weak corrections to $A_{\rm FB}$ become large for high
di-lepton invariant masses. The forward-backward asymmetry at the 
Tevatron for $m(l^+l^-)> 200$~GeV is shown in
Fig.~\ref{fig:fig11}. The QED corrections gradually increase in size
with increasing invariant masses. For $m(l^+l^-)=200$~GeV
($m(l^+l^-)=600$~GeV) they decrease $A_{\rm FB}$ by 0.007 (0.012), i.e. by 
about 1.2\% (2.0\%). The non-factorizable weak corrections further
reduce $A_{\rm FB}$ for the range of masses shown and increase steadily in
size with $m(l^+l^-)$. For $m(l^+l^-)>600$~GeV they are larger than the
QED corrections. The difference in shape between the forward-backward
asymmetry for electrons and muons in the final state at large di-lepton
invariant masses is due to the different momentum resolution for the two
particles. For high energy electrons, the resolution $\sigma/E$ is becoming
independent of the momentum. For high energy muons, on the other hand, the
resolution is proportional to the momentum of the particle,
and $\sigma/p$ becomes of ${\cal O}(1)$ for momenta in the several hundred GeV
region. The momentum of a high energy muon thus is easily mis-measured by
a factor~2 or more, thus modifying $A_{\rm FB}$. 

\begin{figure}
\phantom{x}
\vskip 14.cm
\includegraphics{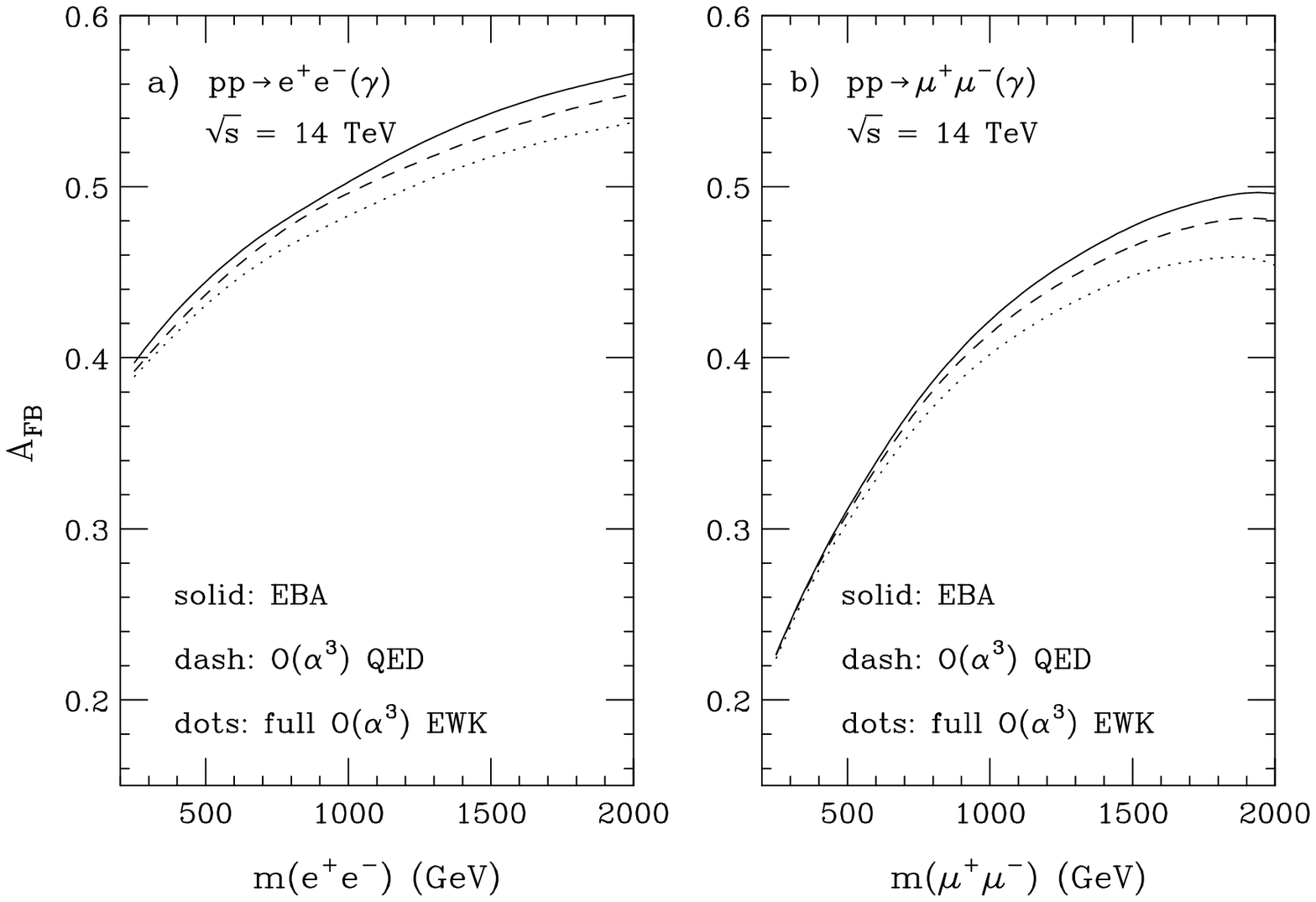}
\caption{The forward-backward asymmetry as a function of $m(l^+l^-)$ for
a) $pp\to e^+e^-(\gamma)$ and b) $pp\to \mu^+\mu^-(\gamma)$ at
$\sqrt{s}=14$~TeV. Shown are the asymmetry in the EBA (solid lines),
including pure QED corrections in addition to those corrections which
are part of the EBA (dashed lines), and the asymmetry 
taking the complete set of ${\cal O}(\alpha)$
electroweak corrections and the ${\cal O}(g^4m_t^2/M_W^2)$ corrections
into account (dotted lines). The cuts and lepton identification requirements
imposed are described in Sec.~\ref{sec:prelim} and~\ref{sec:afb}. 
}
\label{fig:fig12}
\end{figure}

The forward-backward asymmetry for di-lepton invariant masses between
200~GeV and 2~TeV at the LHC is shown in Fig.~\ref{fig:fig12}. For a
$l^+l^-$ invariant mass of 2~TeV, the weak corrections 
are about a factor two larger than the QED corrections. The full
${\cal O}(\alpha)$ electroweak corrections reduce $A_{\rm FB}$ by 0.025
(0.04) for electrons (muons), i.e. by about 5\% (8\%). While the 
forward-backward asymmetry in the high
di-lepton invariant mass region at the Tevatron is nearly constant, it
increases significantly with $m(l^+l^-)$ at the LHC. 
For growing di-lepton invariant masses, the
average fraction $x$ of the proton momentum carried by the quarks
increases, and, as discussed before, this leads to larger values of 
$A_{\rm FB}$.  The large
difference in the magnitude of $A_{\rm FB}$ for electrons and muons is due
to the fact that one of the electrons may have rapidity up to
$|\eta|=4.9$, whereas both muons have to be in the range
$|\eta(\mu)|<2.5$. 

While the non-factorizable weak corrections to the forward-backward
asymmetry increase in magnitude with energy, they 
are significantly smaller than for the di-lepton invariant mass
distribution. This is mostly due to an accidental cancellation of the
box diagrams and the $W$ contributions to the vertex
corrections~\cite{cc}. 

\section{Conclusions}\label{sec:concl}
Drell-Yan production in hadronic collisions is an important 
process. With the anticipated large data sets from Run~II of the
Tevatron and the LHC, it is vital to understand higher order QCD and
electroweak radiative corrections. In this paper we have presented a
calculation of the ${\cal O}(\alpha)$ corrections to 
$p\,p\hskip-7pt\hbox{$^{^{(\!-\!)
}}$} \rightarrow \gamma,\, Z \rightarrow l^+ l^-$ based on the
complete set of one-loop Feynman diagrams contributing to di-lepton
production. In addition, our calculation takes into account the effects
of the ${\cal O}(g^4m^2_t/M_W^2)$ corrections on 
$\sin^2\theta_{\rm eff}^l$ and $M_W$. The calculation is based on a
combination of analytic and Monte Carlo integration techniques. Lepton
mass effects are included in the approximation where only mass singular 
terms originating from the collinear singularity associated with final
state photon radiation are retained. The ultraviolet
divergences associated with the virtual corrections are regularized
using dimensional regularization and the {\sc on-shell} renormalization 
scheme~\cite{bo86}. A previous calculation took the 
${\cal O}(\alpha)$ QED
corrections into account~\cite{Baur:1998wa}, but ignored the effects of
non-universal weak corrections. 

The electroweak ${\cal O}(\alpha)$ corrections to neutral-current
Drell-Yan production naturally decompose into QED and weak contributions
which are separately gauge invariant. The QED corrections can be further
divided into gauge invariant subsets corresponding to initial and
final-state radiation. The collinear singularities associated with
initial state photon radiation are universal to all orders in
perturbation theory and can be absorbed by a redefinition of the parton
distribution functions. The weak corrections can be written in form of
momentum dependent effective vector and axial-vector couplings, and
contributions from $WW$ and $ZZ$ box diagrams. 

Since the phenomenological implications of the QED corrections were
discussed in an earlier paper~\cite{Baur:1998wa}, we concentrated on the
effect of the weak 
corrections on the di-lepton invariant mass distribution and the
forward-backward asymmetry. The weak corrections were found to enhance
the cross section in the $Z$ peak region by about 1\%. In contrast, QED
corrections reduce the cross section in this region by several percent
with the exact amount depending on the flavor of the charged lepton in
the final state, and the lepton identification requirements
imposed. Since the weak corrections are non-uniform in the vicinity of
the $Z$ peak, they are expected to shift the $Z$ boson mass extracted
from data by several MeV. Comparison with the expected statistical
precision of about 0.2\% (0.05\%) for the cross section in the $Z$ peak
region for 2~fb$^{-1}$ (10~fb$^{-1}$) at the
Tevatron (LHC) shows that it will be necessary to take the weak 
corrections into account when one uses observables such as the $Z$ 
boson cross section or the $W/Z$ cross section ratio to confront data
and SM predictions, or when calibrating detector components using $Z$
data. 

The non-universal weak corrections were found to have only a small
effect on the 
forward-backward asymmetry for $m(l^+l^-)<200$~GeV. However, threshold
effects associated with the $WW$ box diagrams lead a characteristic 
peak in $A_{\rm FB}$ at $m(l^+l^-)\approx 160$~GeV. Unfortunately, the size
of this footprint of non-factorizable weak corrections is small, and
thus will be difficult to observe both at the Tevatron and the LHC.

In the $Z$ peak region, the forward-backward asymmetry provides a tool
to measure the effective weak mixing angle. Electroweak
corrections were found to shift $\sin^2\theta^l_{\rm eff}$ by an amount
similar to or larger than the uncertainty expected in future Tevatron
and LHC experiments, and thus cannot be neglected when extracting the
effective weak mixing angle from data. The non-universal weak
corrections, however, contribute only $2.5-10\%$ to the shift. 

The non-factorizable weak corrections to the di-lepton invariant mass
distribution and the forward-backward asymmetry were found to increase
rapidly with $m(l^+l^-)$. This is due to the presence of Sudakov-like
electroweak logarithms of the form $\ln(m(l^+l^-)/M_V)$ ($V=W,\,
Z$). While these corrections are of moderate size
(typically a few percent) for di-lepton invariant masses accessible  
at the Tevatron, they become very large for masses in the TeV region
which play an important role in new physics searches at the
LHC. For $m(l^+l^-)=2$~TeV, the nonfactorizable weak  corrections
reduce the cross section at
the LHC by about 12\%. When QED corrections are taken into account, the
differential cross section may be reduced by as much as 40\% (see
Fig.~\ref{fig:fig8}). 
The strong increase of the non-factorizable weak corrections with the
di-lepton invariant mass requires that these corrections be
resummed. Several calculations of the resummed cross section for 
fermion pair production in $e^+e^-$ collisions have been carried out
recently~\cite{resum}, however, no such calculation exists
yet for di-lepton production in hadronic collisions. 

While the electroweak radiative corrections to the di-lepton invariant 
mass distribution at high values of $m(l^+l^-)$ are very large, they 
were found to be numerically less important in the forward-backward 
asymmetry. 
\acknowledgements
We would like to thank S.~Dittmaier, S.~Eno, H.~Frisch, B.~Knuteson, 
K.~Sliwa and J.~Womersley for stimulating discussions. 
We also thank T. Hahn for cross checking our results for the
weak one-loop corrections with {\it FeynArts}, {\it FormCalc} 
and {\it LoopTools}~\cite{hahn}. One of us (U.B.) is grateful to the 
Fermilab Theory
Group, where part of this work was carried out, for its generous 
hospitality. This work has been supported in part by Department of Energy 
contract No.~DE-FG02-91ER40685 and NSF grants PHY-9600155 and PHY-9970703. 

%
%-----------------------------------------------------------------------
%   Appendix
%-----------------------------------------------------------------------
\newpage
\begin{appendix}
%------------------------------------
% The Z width
%------------------------------------
\section{The Z decay width}\label{app:zdecay}
\setcounter{equation}{0}\setcounter{footnote}{0}

The total $Z$ decay width $\Gamma_Z$ is obtained from the sum over
the partial decay widths into fermion pairs as follows 
\begin{equation}\label{eq:totz}
\Gamma_Z = \sum_{f \neq t} \Gamma_{f\bar{f}} \; .
\end{equation}
At lowest order in perturbation theory the
partial decay widths are given by
\begin{equation}\label{eq:partial0}
\Gamma_{f\bar{f}}^{(0)} =  N_f^C \, \Gamma_0 \, 
       \sqrt{1 - 4 \mu_f} \, \Bigl[ (1 + 2\,\mu_f)\, v_f^2 
        + (1 - 4\,\mu_f)\, a_f^2 \Bigr]
\end{equation}
with the color factor $N_f^C=1,3$, $f=l,q$ ,
\begin{equation}
\Gamma_0 = \frac{\alpha\, M_Z}{3} \qquad \mbox{and} \qquad
\mu_f = \frac{m_f^2}{M_Z^2} \; .
\end{equation}
The fermionic partial decay widths including electroweak and QCD 
radiative corrections can be expressed in terms of 
the effective coupling constants $g_f^{Z,V}, g_f^{Z,A}$ and the
$Z$ wave function renormalization contribution $\hat \Pi^Z$
of Eqs.~(\ref{eq:effcoup}) and~(\ref{eq:piz}):
\begin{eqnarray}\label{eq:partial1}
\Gamma_{f\bar{f}}^{(0+1)} &= & N_f^C \, \Gamma_0 \, 
       \frac{\sqrt{1 - 4 \mu_f}}{1+{\cal R}e \hat \Pi^Z(M_Z^2)} \, 
\left[ (1 + 2\,\mu_f)\, \left\vert g_V^{Z,f}(M_Z^2)\right\vert^2 
        + (1 - 4\,\mu_f)\, \left\vert g_A^{Z,f}(M_Z^2)\right\vert^2 \right]
\nonumber \\[2.mm]
& \times &
(1+\delta^f_{QED})\,\left(1+\frac{N^f_C-1}{2}\,\delta_{QCD}\right)~. 
\end{eqnarray}
The QED corrections,
\begin{equation}
\delta^f_{QED} = \frac{3\, \alpha\, Q_f^2}{4\,\pi}~,
\end{equation}
are at most $0.17\%$ of the lowest-order decay width.
The QCD corrections for massless hadronic final
states have been calculated in \cite{Kataev,Kataev:1992dg} and can be
parametrized in the form ($\alpha_s\equiv \alpha_s(M_Z^2)$)
\begin{equation}
\delta_{QCD} = \left( \frac{\alpha_s}{\pi} \right)
               + 1.405 \left( \frac{\alpha_s}{\pi} \right)^2
               - 12.8  \left( \frac{\alpha_s}{\pi} \right)^3
               - \frac{Q_f^2}{4}\, \frac{\alpha\, \alpha_s}{\pi^2}~.
\end{equation}
Here, the ${\cal O}(\alpha\, \alpha_s)$ term has also been added
although it is not a pure QCD contribution.

For $b$-quarks and $\tau$ leptons it is important to take into account
mass effects in the calculation of the electroweak and QCD corrections. 
From a comparison with a calculation for massive external
fermions~\cite{Cornet:1994qa} one finds
\begin{equation}
\Gamma_{b \bar b}^{(0+1)} = \Gamma_{b \bar b}^{(0+1)}(\mu_b=0) - 
0.0088 \; {\rm GeV} \, , \quad 
\Gamma_{\tau \tau}^{(0+1)} = \Gamma_{\tau \tau}^{(0+1)}(\mu_{\tau}=0) 
- 0.00018 \; {\rm GeV} \, .
\end{equation}
%

%------------------------------------
% Renormalized self energies and form factors
%------------------------------------
\section{Renormalized self energies and form factors}\label{app:sv}
\setcounter{equation}{0}\setcounter{footnote}{0}

The renormalized self energies $\hat{\Sigma}^X(q^2)$ ($X=\gamma,Z,\gamma
Z$) of the neutral vector bosons are given by
\begin{eqnarray}\label{eq:rensigg}
\hat{\Sigma}^{\gamma}(q^2)& = & \Sigma^{\gamma}(q^2) - q^2\,
\Pi^{\gamma}(0), \\[3.mm] \label{eq:rensigz}
\hat{\Sigma}^{Z}(q^2) &= &\Sigma^{Z}(q^2) 
- {\cal R}e \Sigma^{Z}(M_Z^2)+
(q^2-M_Z^2) \left[\frac{c_w^2-s_w^2}{s_w^2} \left( \frac{\delta M_Z^2}{M_Z^2}-
\frac{\delta M_W^2}{M_W^2} \right. \right.
\nonumber \\[2.mm]
&-& \left. \left. 2 \frac{s_w}{c_w} 
\frac{\Sigma^{\gamma Z}(0)}{M_Z^2} \right)-\Pi^{\gamma}(0)  \right]\, ,
\\[3.mm] \label{eq:rensiggz}
\hat{\Sigma}^{\gamma Z}(q^2) &=& \Sigma^{\gamma Z}(q^2) 
       - \Sigma^{\gamma Z}(0) 
       - q^2 \frac{c_w}{s_w} \left[ \frac{\delta M_Z^2}{M_Z^2}-
         \frac{\delta M_W^2}{M_W^2} -
         2 \frac{s_w}{c_w} \frac{\Sigma^{\gamma Z}(0)}{M_Z^2} \right]\, ,
\end{eqnarray}
with $\Pi^{\gamma}(0)=(\partial \Sigma^{\gamma}/ \partial q^2)|_{q^2=0}$ 
and the mass renormalization constants
\begin{equation}\label{eq:onmass}
\delta M_Z^2 = {\cal R}e \left(\Sigma^{Z}(M_Z^2)-\frac{[\hat 
\Sigma^{\gamma Z}(M_Z^2)]^2}{M_Z^2+\hat \Sigma^{\gamma}(M_Z^2)} \right), \quad
\delta M_W^2 = {\cal R}e \Sigma^{W}(M_W^2),
\end{equation}
where $\delta M_Z^2$ is calculated via iteration.
$\Sigma^{X}(q^2)\; (X=\gamma,Z,\gamma Z,W)$ denote the unrenormalized 
self energies as the transverse coefficients in the expansion
\begin{equation}
\Sigma^{X}_{\mu \nu}(q^2) = -g_{\mu\nu} \Sigma^{X}(q^2)
    +\frac{q_\mu q_\nu}{q^2} \left[ \Sigma^{X}(q^2)
    -\Sigma^{X}_L(q^2) \right] \; .
\end{equation}
The terms proportional to $q_\mu q_\nu$ yield contributions
proportional to $m_f^2$ in the 
{\sc on-shell} amplitudes and hence vanish in the limit $m_f \to 0$.
Explicit expressions for the unrenormalized vector boson self energies
$\Sigma^X$ ($X=\gamma,Z,\gamma Z, W$) and the renormalized form factors 
$F_V^{(Z,\gamma),f}, G_A^{(Z,\gamma),f}$ are provided in 
Appendix~B and~C.1 of Ref.~\cite{ho90}. 

Higher-order (irreducible) corrections associated with the $\rho$ parameter 
can be taken into account by performing the replacement
\begin{equation}
\frac{\delta M_Z^2}{M_Z^2}-\frac{\delta M_W^2}{M_W^2} 
\to \frac{\delta M_Z^2}{M_Z^2}-\frac{\delta M_W^2}{M_W^2}+\Delta\rho^{HO}
\end{equation}
in Eqs.~(\ref{eq:rensigz}) and~(\ref{eq:rensiggz}), with
\begin{equation}
\Delta\rho^{HO}=3\,\frac{G_{\mu} m_t^2}{8 \pi^2 \sqrt{2}} \; 
\left[\frac{G_{\mu} m_t^2}{8 \pi^2 \sqrt{2}} \;  
\Delta\rho^{(2)}(m_t^2/M_H^2)+c_1 \; \frac{\alpha_s(m_t^2)}{\pi}+c_2 
\left(\frac{\alpha_s(m_t^2)}{\pi}\right)^2 \right] \; .
\end{equation}
The coefficients $c_1$ and $c_2$ describe the
first and second-order QCD corrections to the leading $G_{\mu} m_t^2$ 
contribution to the $\rho$ parameter,
calculated in~\cite{Djouadi:1987gn} and~\cite{Avdeev:1994db}, respectively. 
Their explicit expressions can be found in Ref.~\cite{yellowbook95} 
(Eqs.~(83,84)). $\alpha_s(m_t^2)$ is given by the following relation:
\begin{equation}
\alpha_s(m_t^2) = \frac{12 \pi}{23}  \; 
\left[\ln\left(\displaystyle \frac{m_t^2}{M_Z^2}\right) +
\displaystyle \frac{12 \pi}{23 \alpha_s(M_Z^2)} \right]^{-1} \; .
\end{equation}
The function $\Delta\rho^{(2)}(m_t^2/M_H^2)$ describes the leading
two-loop electroweak corrections to the $\rho$ parameter and can be
found in Ref.~\cite{Barbieri:1993dq}. 

%------------------------------------
% The box contribution
%------------------------------------
\section{The box contribution}\label{app:box}
\setcounter{equation}{0}\setcounter{footnote}{0}

Two different topologies of weak box diagrams contribute to 
the process $q\bar{q} \to \gamma, Z \to l^+ l^-$ which we denote as 
'direct' and 'crossed' box diagrams. 
The corresponding contributions to the one-loop matrix element are
labeled by $D$ and $C$, accordingly.
Since all fermion mass effects, except the logarithmically divergent
terms associated with the final state collinear divergences, are neglected,
Higgs boson exchange does not contribute, and we only have to consider
box diagrams involving $Z$ and $W^\pm$ exchange. For down (up) type quarks, 
the crossed (direct) box diagram with $W^\pm$ exchange does not
contribute. 

The contribution of the box diagrams to the differential cross section
$d\hat \sigma_{{\rm box}}$ of Eq.~(\ref{eq:xsecweak}) 
thus can be decomposed in the following way
\begin{equation}\label{eq:sbox}
d\hat \sigma_{{\rm box}} = dP_{2f}\;\frac{32 \pi \alpha^3}{3} 
\, {\cal R}e \!\sum_{V=Z,W}  \left[ B_D(\hat s,\hat t,M_V) + B_C(\hat
s,\hat u,M_V) \right]\, ,
\end{equation}
with
\begin{eqnarray}
B_D(\hat s,\hat t,M_V) &=& \kappa^+_V \, (\hat s+\hat t)^2 \, 
[2 D_2^0+\hat t \, (D_1^1+D_1^2+D_1^3+D_2^2+D_2^{23}+D_2^{12})]
\nonumber \\
&+& \kappa^-_V \, \hat t^2 \, 
[8 D_2^0+\hat t \, (D_1^1+2 D_1^2+D_1^3+2 (D_2^2+D_2^{23}+D_2^{12}))
-2 \hat s D_2^{13}]\, ,
\nonumber \\
B_C(\hat s,\hat u,M_V) &=& - B_D(\hat s,\hat t,M_V) \quad  \mbox{ with} \quad 
\hat t \leftrightarrow \hat u, \, \kappa^+_V \leftrightarrow \kappa^-_V , 
\end{eqnarray}
where the four-point functions are denoted by 
$D_i^j=D_i^j(\hat t,0,M_V,0,M_V)$ and
\begin{eqnarray}
\kappa^+_W &=& \frac{1}{8 s_w^4} \left[(v_l+a_l)(v_q+a_q)
                 \frac{(\hat s-M_Z^2)}{|\hat s-M_c^2|^2}
                  + \frac{Q_l Q_q}{\hat s} \right]\, ,
\nonumber \\[2.mm]
\kappa^-_W &=& 0\, ,
\nonumber \\
\kappa^+_Z &=& \left[(v_l^3+3 v_l a_l^2)(v_q^3+3 v_q a_q^2)
               +(a_l^3+3 a_l v_l^2)(a_q^3+3 a_q v_q^2)\right]  
               \frac{(\hat s-M_Z^2)}{|\hat s-M_c^2|^2}
\nonumber \\[1.mm]
& + & \frac{Q_l Q_q}{\hat s} 
\left[(v_l^2+a_l^2)(v_q^2+a_q^2)+4 v_l v_q a_l a_q\right]\, ,
\nonumber \\[2.mm]
\kappa^-_Z &=& \left[(v_l^3+3 v_l a_l^2)(v_q^3+3 v_q a_q^2)
               -(a_l^3+3 a_l v_l^2)(a_q^3+3 a_q v_q^2)\right]  
               \frac{(\hat s-M_Z^2)}{|\hat s-M_c^2|^2}
\nonumber \\[1.mm]
& + & \frac{Q_l Q_q}{\hat s} \left[(v_l^2+a_l^2)(v_q^2+a_q^2)-4 v_l v_q
a_l a_q\right]\, ,
\end{eqnarray}
with the vector and axial vector couplings, $v_f, a_f (f=l,q)$, 
of Eq.~(\ref{eq:coup}) and $M_c$ being the complex $Z$ boson mass.
The explicit decomposition of the vectorial and tensorial four point 
functions
\begin{eqnarray}\label{eq:dmunuint}
\lefteqn{\frac{i}{16 \pi^2}\, D_{\mu,\mu \nu}(\hat t,m_l,M_V,m_q,M_V) = }
\nonumber \\[3.mm]
& & \int \frac{d^4 k}{(2 \pi)^4}
\frac{k_{\mu},k_{\mu} k_{\nu}}{(k^2-m_l^2)((k-k_+)^2-M_V^2)((k+p-k_+)^2-m_q^2)
((k+k_-)^2-M_V^2)} ~,
\end{eqnarray}
with
\begin{eqnarray}\label{eq:dmunu}
D_{\mu}&=&-k_{+ \mu} D_1^1 + (p-k_+)_\mu D_1^2+k_{- \mu} D_1^3  \, ,
\nonumber \\
D_{\mu \nu}&=&k_{+ \mu} k_{+ \nu} D_2^1 + (p-k_+)_\mu (p-k_+)_\nu D_2^2
+k_{- \mu} k_{- \nu} D_2^3 + g_{\mu \nu} D_2^0 
\nonumber \\
&-& [ k_{+ \mu} (p-k_+)_\nu + k_{+ \nu} (p-k_+)_\mu] D_2^{12}
-( k_{+ \mu} k_{- \nu} +  k_{+ \nu} k_{- \mu}) D_2^{13}
\nonumber \\
&-& [ k_{+ \mu} (p-k_+)_\nu + k_{+ \nu} (p-k_+)_\mu] D_2^{12}
+[k_{- \mu} (p-k_+)_\nu + k_{-\nu} (p-k_+)_\mu] D_2^{23}\, ,
\end{eqnarray}
can be found in Ref.~\cite{Sack:1987qf}. The expressions for 
the crossed box diagram, $D_{\mu,\mu \nu}(\hat u,m_l,M_V,m_q,M_V)$ 
can be obtained by replacing $p \to \bar p$
in Eqs.~(\ref{eq:dmunuint}) and~(\ref{eq:dmunu}).

\end{appendix}
%
%%% References %%%
%

%

\newpage
%
%%%%%%%%%%%%%%%%%%%%%%%%% TABLES %%%%%%%%%%%%%%%%%%%%%%%%%%%%%%%%%%%%%%%%
\widetext

\begin{table}
\caption{The cross section ratios $K^{EW}=\sigma^{{\cal O}(\alpha^3)}
/\sigma^{\rm EBA}$ and $K^{\rm QED}=\sigma^{\rm QED}/\sigma^{\rm 
EBA}$ for $p\bar p\to l^+l^-(\gamma)$ ($l=e,\,\mu$) at 
$\protect{\sqrt{s}=2}$~TeV
with \protect{$75~{\rm GeV}<m(l^+l^-)<105$}~GeV. Shown are the 
predictions with and without taking cuts and lepton identification
requirements into account.}
\label{tab:one}
\vskip 3.mm
\begin{tabular}{ccccc}
 & \multicolumn{2}{c}{with lepton id.} & \multicolumn{2}{c}{without
lepton id.} \\
 & \multicolumn{2}{c}{requirements} & \multicolumn{2}{c}{requirements} 
\\
channel & $K^{EW}$ & $K^{\rm QED}$ & $K^{EW}$ & $K^{\rm QED}$ \\
\tableline
$p\bar p\to e^+e^-(\gamma)$ & 0.988 & 0.978 & 0.949 & 0.939 \\
$p\bar p\to\mu^+\mu^-(\gamma)$ & 0.936 & 0.926 & 0.981 & 0.971
\end{tabular}
\end{table}

\vskip 1.cm
\begin{table}
\caption{The coefficients $a$ and $b$ defined in Eq.~(\ref{EQ:AFB}) in
the EBA and the shifts introduced by QED and weak corrections for the
integrated forward-backward asymmetry in the region $75~{\rm
GeV}<m(l^+l^-)<105$~GeV at the Tevatron and the LHC. The values listed
for the Tevatron are obtained without imposing any cuts or lepton
identification requirements. At the LHC, the cuts discussed in
Sec.~\ref{sec:afb} have been imposed, together with the lepton
identification requirements listed in Sec.~\ref{sec:prelim}.}
\label{tab:two}
\vskip 3.mm
\begin{tabular}{c|ccc|ccc}
\multicolumn{7}{c}{a) Tevatron}\\
final state & $a^{\rm EBA}$ & $\Delta a^{\rm QED}$ & $\Delta a^{\rm weak}$ &
$b^{\rm EBA}$ & $\Delta b^{\rm QED}$ & $\Delta b^{\rm weak}$ \\
\tableline
$e^+e^-(\gamma)$ & 0.24585 & 0.00221 & $-0.00016$ & 3.408 & $-0.408$ &
0.026 \\
$\mu^+\mu^-(\gamma)$ & 0.24585 & 0.00094 & $-0.00001$ & 3.408 & $-0.171$ 
& 0 \\
\tableline
\multicolumn{7}{c}{b) LHC}\\
final state & $a^{\rm EBA}$ & $\Delta a^{\rm QED}$ & $\Delta a^{\rm weak}$ &
$b^{\rm EBA}$ & $\Delta b^{\rm QED}$ & $\Delta b^{\rm weak}$ \\
\tableline
$e^+e^-(\gamma)$ & 0.24797 & 0.00284 & $-0.00020$ & 1.618 & $-0.276$ &
0.013 \\
$\mu^+\mu^-(\gamma)$ & 0.25072 & $-0.00012$ & 0.00037 & 0.724 & $-0.050$ 
& 0.013
\end{tabular}
\end{table}

\newpage
%
%%%%%%%%%%%%%%%%%%%%%% FIGURE CAPTIONS %%%%%%%%%%%%%%%%%%%%%%%%%%%%%%%%%%
%
% FIG. 1
%

\vskip 3.cm

\newpage

\newpage

\newpage

\newpage

\newpage

\newpage

\newpage

\newpage

\newpage

\newpage

\end{document}